\documentclass[journal,twoside,10pt,twocolumn]{IEEEtran}
\usepackage{graphicx,amssymb,amsmath,amsthm,psfrag,stfloats,color,dsfont,subfigure,listings,epstopdf,cite,bm}
\usepackage{algorithm,algorithmic}
\usepackage[skip=-3pt]{caption}

\begin{document}

\title{\fontsize{19.5pt}{19}\selectfont Performance Analysis of STAR-RIS-Assisted NOMA Wireless Systems with Realistic Indoor Outdoor THz Channel Models}

\author{Ngoc Phuc Le, \IEEEmembership{Member, IEEE}, and Mohamed-Slim Alouini, \IEEEmembership{Fellow, IEEE}
\thanks{Copyright \textcopyright~20xx IEEE. Personal use of this material is permitted. However, permission to use this material for any other purposes must be obtained from the IEEE by sending a request to pubs-permissions@ieee.org.}
\thanks{The authors are with the Division of Computer, Electrical and Mathematical Sciences and Engineering, King Abdullah University of Science and Technology (KAUST), Thuwal 23955-6900, Saudi Arabia (e-mails: phucle.ngoc@kaust.edu.sa;slim.alouini@kaust.edu.sa).}

\vspace*{-0.5cm}}
\maketitle

\begin{abstract}
In this paper, a simultaneously transmitting and reflecting reconfigurable intelligent surface (STAR-RIS)-aided downlink non-orthogonal multiple access (NOMA) Terahertz (THz) wireless system is proposed for indoor and outdoor transmissions. We consider a near-field communication scenario where an access-point (AP) is deployed near a STAR-RIS panel. For links from the STAR-RIS to users, $\alpha-\mu$ distribution is adopted for the indoor small-scale fading channels, whereas the outdoor channels are based on Gaussian mixture or mixture of gamma, which follows the recent practical measurement reports. To facilitate performance analysis, we derive exact expressions of a probability density function (PDF) and a cumulative distribution function (CDF) of a weighted sum of $\alpha-\mu$ variates. Approximate PDF and CDF expressions of a weighted sum of Gaussian mixture variates are derived as well. Based on these results, closed-form expressions of the outage probability and the ergodic capacity, together with their asymptotic formulas at high signal-to-noise ratio (SNR), are obtained. Moreover, we analyze the capacity of the THz system at the low SNR regime. Impacts of hardware impairments and STAR-RIS protocols (i.e., energy splitting and mode-switching) on the system performance are evaluated. All developed analytical results are validated and demonstrated via numerical simulations.         
\end{abstract}


\makeatletter
\long\def\@makecaption#1#2{\ifx\@captype\@IEEEtablestring%
\footnotesize\begin{center}{\normalfont\footnotesize #1}\\
{\normalfont\footnotesize\scshape #2}\end{center}%
\@IEEEtablecaptionsepspace
\else
\@IEEEfigurecaptionsepspace
\setbox\@tempboxa\hbox{\normalfont\footnotesize {#1.}~~ #2}%
\ifdim \wd\@tempboxa >\hsize%
\setbox\@tempboxa\hbox{\normalfont\footnotesize {#1.}~~ }%
\parbox[t]{\hsize}{\normalfont\footnotesize \noindent\unhbox\@tempboxa#2}%
\else
\hbox to\hsize{\normalfont\footnotesize\hfil\box\@tempboxa\hfil}\fi\fi}
\makeatother

\section{Introduction}
\IEEEPARstart{F}ramework and objectives of the future developments for International Mobile Telecommunications-2030 and Beyond (IMT-2030) has just released recently in \cite{ITU}. Accordingly, IMT-2030 is expected to provide enhanced and new capabilities compared to IMT-2020, including immersive communication, hyper reliable and low-latency communication, massive communication, integrated sensing and communication (ISAC), articial intelligent (AI) and communication, and ubiquitous connectivity. Also, emerging technology enablers to fulfill these usage scenarios were identified, such as reconfigurable intelligent surfaces (RIS), non-orthogonal multiple access (NOMA), and frequency bands over 100 GHz (i.e., millimeter wave and Terahertz (THz)) \cite{ITU}-\hspace{1pt}\cite{Dan}.

Requirements of the next generation IMT use cases, e.g., extremely high data rate, low latency, and highly precise positioning, etc., can be supported by THz communications thanks to vast bandwidth resources available \cite{Ela}-\hspace{1pt}\cite{Aky}. However, the deployment of THz incurs a well-known issue of high path loss. Moreover, THz signals are more vulnerable to blockages. These challenges can be overcome with the help of RIS. In particular, RIS comprises of several unit-cells whose properties, such as reflection, refraction, and absorption, can be controlled to construct an intelligent and programmable radio environment \cite{Bas}-\hspace{1pt}\cite{Wu}. Therefore, RIS can improve transmission reliability and achieve higher spectrum efficiency. Additionally, RIS creates virtual line-of-sight (LoS) links, which is helpful to tackle blockage issues and expand network coverage.       

\subsection{Review of Related Works} 
\subsubsection{RIS for THz Wireless Systems} 
To exploit the potential benefits of RIS and THz technologies, many research works have recently studied RIS-aided THz systems \cite{Che}-\hspace{1pt}\cite{Cha}. Several research issues associated with RIS-aided THz systems under various usage cases have been considered. Generally, the aim is to improve performance in terms of outage probability, capacity, and energy efficiency. Literature review of RIS-aided THz wireless systems can be found in \cite{Che}-\hspace{1pt}\cite{Le}. In what follows, we focus on highlighting and updating the readers with key research related to our proposed system in this study.    
 
Firstly, one of the most essential research problems on RIS-aided THz is phase-shifts optimization or joint optimization of phase-shifts with other system parameters to enhance the system performance. In \cite{Pan}, conventional convex optimization approach is used to jointly optimize beamforming matrix at the transmitter and phase-shifts at RIS for improved sum-rate. Meanwhile, machine-learning based approach was adopted in \cite{Hua}. In addition, optimizing parameters in RIS-aided THz systems to support heterogeneous data-rate services or ISAC were investigated in \cite{Zar}-\hspace{1pt}\cite{He}.   

Secondly, several types of RIS schemes were proposed for THz systems. In particular, \cite{Hua} considered multi-hop (i.e., cascaded) RIS-aided THz, whereas \cite{Huo} studied distributed RIS-aided THz systems. RIS for multi-input multi-output (MIMO) systems were investigated in \cite{Pan} and \cite{Nin}. In addition, RIS-aided THz was considered for non-terrestrial networks, e.g., inter-satellite links of low-Earth orbit (LEO) networks \cite{Tek} or unmanned aerial vehicle (UAV)-based system \cite{Pan1}. It is worth noting that these studies considered only passive RIS for THz systems. To deal with impacts of the multiplicative fading effect, which is inherently with passive RIS deployments, we proposed in \cite{Le1} hybrid passive-active RIS for THz systems. Specifically, we analyzed the outage probability and capacity performance, and shown that hybrid RIS-aided THz can achieve higher energy-efficiency than its counterparts.

Thirdly, several authors have studied RIS-aided THz systems with small-scale fading. This is motivated by the fact that there exists small-scale fading at THz transmissions \cite{Pap}-\hspace{1pt}\cite{Pap1}. In particular, analytical expressions of the outage probability and ergodic capacity in RIS-aided THz systems with the fluctuating two-ray distribution were derived in \cite{Du}. Also, exact and approximate analysis of RIS-aided THz systems over $\alpha-\mu$ fading were carried out in \cite{Le} and \cite{Cha}, respectively. Note that \cite{Cha} also took into consideration the impact of pointing errors since it is an important issue for two-hop THz transmissions as shown in \cite{Li7}. Additionally, performance of active RIS-assisted mixed radio frequency (RF)-THz relaying systems was studied in \cite{Yin1}. Specifically, the authors derived exact and asymptotic expressions of the outage probability, the average bit-error rate, and the channel capacity.

\subsubsection{STAR-RIS-assisted Wireless Systems}
Conventional RIS is used for reflection of incident signals, and thus it is limited to half-space coverage. Recently, a new RIS architecture, namely simultaneously transmitting and reflecting RIS (STAR-RIS), was devised, that can offer full-space coverage \cite{Xu}. Specifically, the incident wireless signals on a STAR-RIS panel can be transmitted or reflected on both sides of the surface, which enables $360^o$ transmission coverage. Three operating protocols were proposed for STAR-RIS, namely energy splitting (ES), mode switching (MS), and time switching (TS) \cite{Liu}-\hspace{1pt}\cite{Mu}. In the ES protocol, all RIS elements work in both transmission and reflection modes. On the other hand, RIS elements of the MS protocol work only in either a transmission mode or a reflection mode, whereas RIS elements of the TS protocol periodically work in a transmission mode and a reflection mode over different time slots. Also, different hardware models/implementations and channel models for STAR-RIS were reported in \cite{Xu}, \cite{Zha}, and \cite{Xu1}. 

Extensive research efforts have been devoted to STAR-RIS-based wireless systems. One of the most essential research problems in STAR-RIS is beamforming/optimizing phase-shifts. In \cite{Mu}, a joint active beamforming at a transmitter and phase-shift optimizing at a STAR-RIS was formulated and solved for power consumption minimization in both unicast and multicast transmissions. The results shown that STAR-RIS could significantly reduce power consumption compared to conventional RIS. A joint optimization problem for sum-rate maximization in STAR-RIS-assisted NOMA was considered in \cite{Zuo}. For cases of coupled phase-shifts, a generalized optimization framework in STAR-RIS was proposed to maximize the throughput in \cite{Wan0}. Also, a hybrid reinforcement learning approach was proposed in \cite{Zho} to obtain the transmission/reflection coefficients under coupled phase-shift constraints. In addition, deep reinforcement learning was adopted for energy-efficient design in a STAR-RIS-aided NOMA network in \cite{Guo}.
 
Another important research venue in STAR-RIS is performance analysis. In \cite{Zha1}, the authors investigated the performance of a STAR-RIS-aided downlink NOMA system with randomly deployed users. Specifically, they derived closed-form expressions of outage probability and diversity gains under ES, MS, and TS protocols. Performance analysis in terms of outage probability and power scaling law in cases of coupled phase-shifts was performed in \cite{Xu2}. Also, a closed-form expression of the coverage probability in a STAR-RIS assisted massive MIMO system was derived in \cite{Pap3}. In \cite{Yue1}, the authors derived approximate and asymptotic expressions of the outage probability and the ergodic capacity in STAR-RIS NOMA networks in the presence of perfect/imperfect successive interference cancellation (SIC) over Rician fading channels. The system throughput for both delay-tolerant and delay-limited transmission modes was investigated in this work as well. A further study of active STAR-RIS NOMA systems with uniformly distributed paring users was presented in \cite{Yue2}. Also, performance comparison between STAR-RIS NOMA systems with their STAR-RIS OMA counterparts were considered in these two research works. Additionally, performance analysis of downlink STAR-RIS NOMA systems in the presence of residual hardware impairments and imperfect channel state information (CSI) over Rayleigh fading channels was investigated in \cite{Zha2}.

\subsubsection{ Near-Field RIS-THz Wireless System} 
Recently, near-field communication has gained significant attention due to the promises of extremely large-scale MIMO (XL-MIMO)/large RIS and very high frequency transmissions. In contrast to a planar wavefront assumption in conventional far-field communications, near-field exhibits distinctive features, which necessitates developments of novel channel modeling, channel estimation, and beamfocusing \cite{An3}. In \cite{Xu10}, the authors investigated STAR-RIS in near-field communications with a Green’s function method based channel modeling. They proposed three STAR-RIS configuration strategies, namely power splitting, selective element grouping, and random element grouping. The channel gains were also derived for both the pure near-field regime and the hybrid near-field and far-field regime. Meanwhile, near-field wideband beamforming for RIS-assisted MIMO systems was developed for the maximal spectral efficiency in \cite{Wan10}. Specifically, two RIS architectures, namely true time delay-based RIS (TTD-RIS) and virtual subarray-based RIS (SA-RIS), are considered to achieve the frequency-dependent passive beamforming at the RIS. In addition, holographic MIMO based on STAR-RIS NOMA was proposed in \cite{Li6}, where the authors analyze the achievable rate in the presence of hardware impairments.

In the near-field regime, one of the unique features is spatial non-stationarity (SnS) phenomenon, where elements of large-aperture antenna arrays at different spatial positions observe different channel multipath characteristics. In \cite{Yua10}, a realistic yet low-complexity SnS channel modeling framework for massive MIMO systems was proposed. An analytical study of the near-field characteristics of the SNR in XL-MIMO with SnS was presented in \cite{Zhi10}. Also, a channel estimation framework based on multi-task learning in hybrid near-field/far-field STAR-RIS systems with SnS was investigated in \cite{Xia10}.

\subsection{Motivations and Contributions}
To exploit the potential benefits of RIS and THz for $360^o$ coverage with improved reliability and capacity, it is obviously essential to study STAR-RIS-assisted THz systems. To this end, some works have recently considered STAR-RIS-assisted THz systems, e.g., \cite{Wan}-\hspace{1pt}\cite{Mah}. In \cite{Wan}, the authors jointly optimized the hybrid beamforming at a transmitter and the passive beamforming at a STAR-RIS to maximize spectral efficiency and energy efficiency. Also, a joint optimization of transmit beamforming and phase-shifts of beyond-diagonal RIS-THz with a hybrid reflection/transmission mode was investigated in \cite{Mah}. However, these studies assume no small-scale fading for THz transmissions, which is not applicable to several practical scenarios. Moreover, analytical analysis of key performance metrics such as outage probability or capacity was not performed.

In contrast to the existing works mentioned above, the proposed STAR-RIS-assisted THz system model in this study takes into consideration several realistic aspects. First, we adopt $\alpha-\mu$ distribution for the indoor small-scale fading channels, whereas the outdoor channels are based on either Gaussian mixture (GM) or mixture of gamma (MoG). These channel modelings are based on recent practical measurements for THz links reported in \cite{Pap} and \cite{Pap1}, respectively. Second, we consider the presence of residual hardware impairments (HWI) at the transceivers. Third, we investigate the near-field communication scenario. The main contributions of our work are summarized below.
\begin{itemize}
	\item Characterize statistical distribution of a weighted sum of $\alpha-\mu$ variates as well as a weighted sum of Gaussian mixture (GM) variates. In particular, we derive exact expressions of a probability density function (PDF) and a cumulative distribution function (CDF) of a weighted sum of $\alpha-\mu$ variates. Also, accurately approximate PDF and CDF expressions of a weighted sum of GM variates are derived. These new results facilitate analytical description of the end-to-end (e2e) channel distribution in STAR-RIS-assisted THz systems.
	\item Derive closed-form expressions of the outage probability (OP) and the ergodic capacity (EC). We also perform asymptotic analysis of the OP and EC at the high SNR regime. These results provide insights into the performance evaluation of the system.
	\item Analyze the capacity of the system at the low SNR regime, which is interested in THz networks from practical perspectives.
	\item Evaluate impacts of hardware impairments and the STAR-RIS protocols of ES and MS on the performance of the proposed system in all considered scenarios. 
\end{itemize}

\subsection{Organization of the Paper}
The remaining of this paper is organized as follows. In Section II, we describe the STAR-RIS-assisted NOMA THz system model with hardware impairments. Section III develops statistical characterization of the distributions of the e2e channels. Section IV analyzes the OP and EC performance as well as their asymptotic expressions. Simulation results and discussions are provided in Section V. Finally, Section VI concludes the paper.

\section{STAR-RIS-Assisted THz Wireless System Model}
\subsection{System Model}

\begin{figure}[t] 
	\centering{\includegraphics[width=0.45\textwidth]{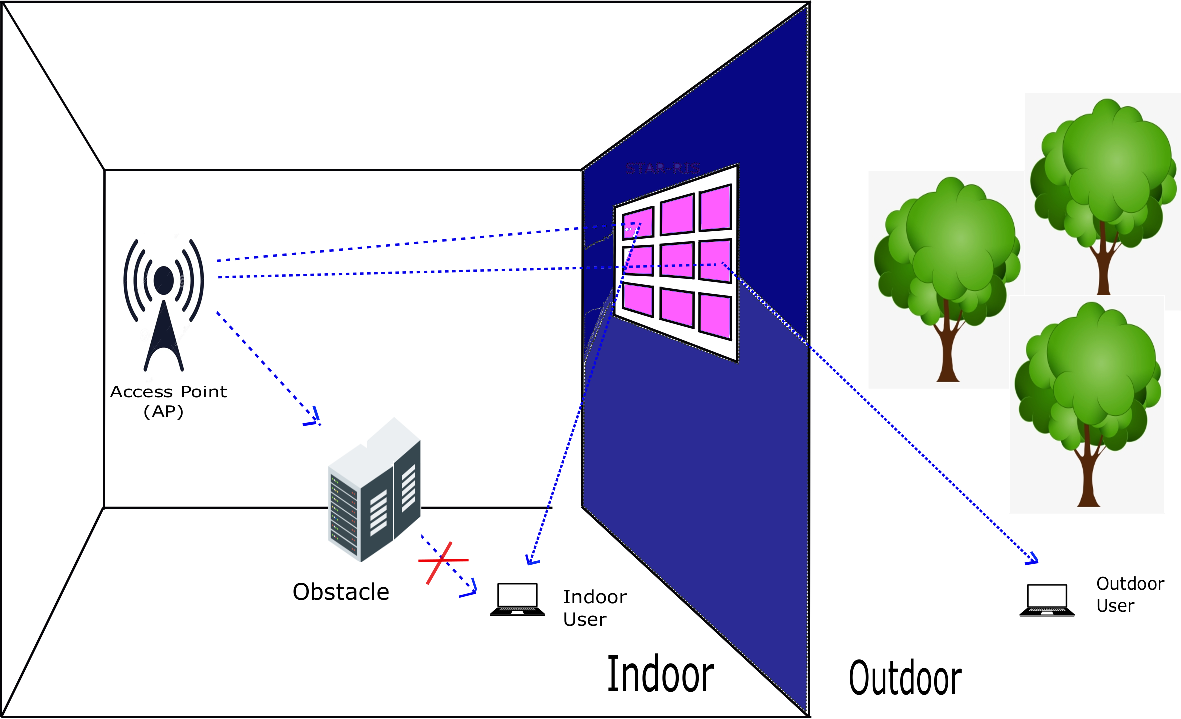}}
	\caption{A STAR-RIS-aided downlink NOMA THz system model.}
	\label{Fig1}
\end{figure}

We study a STAR-RIS-aided downlink NOMA THz wireless system with transceiver's hardware impairments as shown in Fig. 1. A considered system model consists of an access point (AP) and one user located indoor (i.e., on reflective side of RIS), while the other user is outdoor (i.e., on transmissive/refractive side of RIS)\footnote{We consider two NOMA users in this study, which is similar to several existing works on STAR-RIS NOMA in the literature, e.g., \cite{Zha1}-\hspace{1pt}\cite{Li6}. However, it is worth noting that this system model can be easily extended to multi-user scenarios, where users are grouped into two-user clusters, and different clusters are allocated into orthogonal resource blocks. An implementation of STAR-RIS was demonstrated in \cite{Zha20}. Also, real experiments for the outdoor-to-indoor and indoor-to-outdoor transmission via RIS as the glass window of the building were reported in \cite{Kit1} and \cite{Got1}, respectively}. The users are located in the far-field of STAR-RIS, whereas the AP is operated in the near-field due to a short transmission distance from the AP to the STAR-RIS panel \cite{An3}. Also, the AP transmits signals to both indoor and outdoor users via a NOMA principle with the help of the STAR-RIS panel. The coordinate of the AP is $(0,0,-d_0)$, whilst the STAR-RIS panel is centered at the origin on the $xOy$ plane. Here, the STAR-RIS consists of $M$ elements, and the coordinate of the center point of the $m^{th}$ element is denoted by $(x_m,y_m,0), \forall m \in \mathcal{M}$. 

In this work, we consider two STAR-RIS protocols of energy-splitting (ES) and mode-switching (MS) \cite{Xu}, \cite{Mu}\footnote{For brevity, we present an ES protocol in the analysis. However, the results are extended straightforwardly to a MS protocol. Also, performance comparison between the two protocols are provided in Simulation section.}: 

\textit{Energy-splitting (ES)}: For the ES protocol, all STAR-RIS elements simultaneously used for refraction and reflection
operations. Specifically, the energy of the impinging signal on each STAR-RIS element is split into the energies of the
transmitted and reflected signals. Let the amplitude and the phase-shift of the $m^{th}$ RIS element be denoted by $\mathrm{a}_{I,m}$ and $\theta_{I,m}$ for the reflective side, and by $\mathrm{a}_{O,m}$ and $\theta_{O,m}$ for the transmissive side, respectively. Then, the reflection and transmission coefficient matrices of the STAR-RIS are given by $\mathbf{\Theta}_I=\operatorname{diag}\left(\mathrm{a}_{I,1}e^{j\theta_{I,1}}, \mathrm{a}_{I,2}e^{j\theta_{I,2}}, \cdots, \mathrm{a}_{I,M}e^{j\theta_{I,M}}\right)$ and $\mathbf{\Theta}_O=\operatorname{diag}\left(\mathrm{a}_{O,1}e^{j\theta_{O,1}}, \mathrm{a}_{O,2}e^{j\theta_{O,2}}, \cdots, \mathrm{a}_{O,M}e^{j\theta_{O,M}}\right)$, respectively. It is worth noting that $\mathrm{a}_{I,m}^2+\mathrm{a}_{O,m}^2=1, \forall m \in \mathcal{M}$, following an energy conservation law\footnote{In this work, we assume that the STAR-RIS does not impose a power loss or gain amplification. However, our proposed analysis is also applicable to these cases, i.e., $\mathrm{a}_{I,m}^2+\mathrm{a}_{O,m}^2=c_m$, where $c_m>0$.}. Also, for analytical simplicity, we consider the case that the phase-shifts of reflection and transmission are independent, i.e., $\theta_{\chi,m}\in [0,2\pi), \forall m \in \mathcal{M}$, and $\chi \triangleq \{I,O\}$, which is similar to \cite{Xu}, \cite{Mu}, \cite{Zuo}, \cite{Guo}, \cite{Pap3}, \cite{Yue2}-\hspace{1pt}\cite{Zha2}.

\textit{Mode-switching (MS)}: In this protocol, all STAR-RIS elements are partitioned into two groups. One group contains elements that is used for reflection only, while the other contains elements that is for transmission only. This protocol can be regarded as a special case of the ES protocol by constraining the amplitude coefficients $\mathrm{a}_{\chi,m}$ to binary values. In particular, for the reflection group $\mathcal{M}_I$, we have $\mathrm{a}_{I,m}=1, \mathrm{a}_{O,m}=0, \forall m \in \mathcal{M}_I$. Meanwhile, for the transmission group $\mathcal{M}_O$, we have $\mathrm{a}_{I,m}=0, \mathrm{a}_{O,m}=1, \forall m \in \mathcal{M}_O$, where $\mathcal{M}_I \cup \mathcal{M}_O= \mathcal{M}$ and $\mathcal{M}_I \cap \mathcal{M}_O= \emptyset$.
 
We assume that the direct links from the AP and the users are unavailable due to severe blockage, which is similar to many existing works in the literature\footnote{In cases that the direct links exist, the system performance might be improved. Meanwhile, it might impose several challenges in terms of channel estimation, phase coordination between then direct and the RIS-assisted links, and complexity regarding performance analysis and optimization. The readers are referred to \cite{Zhi60} and \cite{Ren10} for more details.}. Thus, the equivalent end-to-end (e2e) channel from the AP to the users can be expressed as
\begin{align} \label{S1}
H_{\chi}=\sum_{m=1}^{M}h_{\chi,m}\mathrm{a}_{\chi,m}e^{j\theta_{\chi,m}}g_{m},
\end{align} 
\noindent where $g_{m}$ is the link from the AP to the $m^{th}$ RIS element, whereas $h_{\chi,m}$ is the link from this element to the user $\chi$. Therefore, the signal received by the user $\chi$ is expressed as
\begin{align} \label{S2}
r_{\chi}=H_{\chi}\left(P_Is_I+P_Os_O+\xi\right)+n_{\chi},
\end{align} 
\noindent where $s_{\chi}$ and $P_{\chi}$ are the transmitted signal and the allocated power associated with the user $\chi$ with $\mathbb{E}\{|s_{\chi}|^2\}=1$. Also, $\xi$ represents the aggregated distortion noise due to non-ideal hardware at the transceiver, and $n_{\chi}$ is the additive white Gaussian noise (AWGN) with zero-mean and variance of $N_0$, i.e., $n_{\chi} \sim \mathcal{CN}(0,N_0)$. Note that the total transmit power is $P=P_I+P_O$, where $P_I=\rho_I P$, $P_O=\rho_O P$, and $\rho_{\chi}$ is the power allocation coefficient, i.e., $\rho_I+\rho_O=1$. Additionally, we assume that $\xi \sim \mathcal{CN}(0,\kappa^2P)$, where $\kappa$ is the level of residual hardware impairments (HWI) \cite{Le}, \cite{Le1}. 

In this system model, downlink NOMA is adopted for transmission. Similar to several downlink NOMA research works, e.g., \cite{Zuo}, \cite{Guo}, \cite{Zha1}, and \cite{Li6}, we assume that indoor user (i.e., near user) has the stronger channel, i.e., $|H_I|>|H_O|$. For user fairness, it is required that more transmit power is allocated to the outdoor user, i.e., $\rho_O>\rho_I$. Consequently, the indoor will implement successive interference cancellation (SIC) for signal detection\footnote{The indoor user is treated as the nearby user in this work to facilitate analysis. In cases the distant user and nearby user are distinguished based on the instantaneous channel coefficients, the order statistics should be employed for performance analysis. This scenario goes beyond the scope of this work and is left for future investigations.}. Specifically, it will first detect the signal of the outdoor user with the signal-to-interference-plus-distortion-and-noise ratio (SIDNR) as
\begin{align} \label{S3}
\gamma_{I \rightarrow O}=\frac{P_O|H_I|^2}{\kappa^2P|H_I|^2+P_I|H_I|^2+N_0},
\end{align}  
\noindent It then subtracts the signal of the outdoor user from its received signal and decodes its information with the SDNR
\begin{align} \label{S4}
\gamma_I=\frac{P_I|H_I|^2}{\kappa^2P|H_I|^2+N_0}.
\end{align} 
\noindent Meanwhile, the outdoor user directly decodes its own signal by treating the signal of the indoor user as interference. As a result, the SIDNR can be evaluated by \cite{Le}
\begin{align} \label{S5}
\gamma_O=\frac{P_O|H_O|^2}{\kappa^2P|H_O|^2+P_I|H_O|^2+N_0}.
\end{align} 

\noindent It is worth noting from Eq. (2) that the presence of HWI adds additive distortion noise to the received signal. As a result, the SDNR and SIDNR expressions in Eq. (4) and Eq. (5) are more complex than the SNR and the SINR counterparts in the ideal system. This will introduce complexity in analysis and impacts the system performance as shown in Section IV and Section V, respectively.

Note that for a downlink STAR-RIS OMA system, data for each user is transmitted over non-overlapping resources (e.g., time slots or frequency bands) to avoid interference between users. Thus, the SDNR of the indoor user and the outdoor user can be express as
\begin{align} \label{S6}
\gamma_{\chi}^{OMA}=\frac{P_{\chi}|H_{\chi}|^2}{\kappa^2P|H_{\chi}|^2+N_0}.
\end{align}

\subsection{Indoor Outdoor THz Channel Models}
For THz transmissions, a channel model will consist of path gain and small-scale fading, i.e., $h=\bar{h}\tilde{h}$, where $\bar{h}$ and $\tilde{h}$ denote the path gain and the small-scale fading coefficient, respectively. The path gain at THz bands is due to propagation gain and molecular absorption gain, which can be expressed as \cite{Le}, \cite{Le1}
\begin{align} \label{Ch1}
\bar{h}=\frac{c\sqrt{G_tG_r}}{4\pi f d}e^{-\frac{1}{2}\varrho(f)d},
\end{align}  
\noindent where $f$ is the carrier frequency, $c$ is the speed of light, $d$ is the transmission distance, $G_t$ and $G_r$ are the antenna gains at the transmit and the receive sides, respectively. Also, $\varrho(f)$ is the molecular absorption coefficient, whose value depends on the operating frequency. More details about the absorption coefficient can be found in \cite{Kok}-\hspace{1pt}\cite{Bab}.

\subsubsection{STAR-RIS to Indoor User Channel Model}
Following a recent experiment report with respect to small-scale fading of THz transmissions, i.e., \cite{Pap}, we adopt $\alpha-\mu$ fading distribution for indoor environments in the far-field scenario. In particular, for a random variable $X$ that follows $\alpha-\mu$ distribution with parameters $(\alpha,\mu,\Omega)$, the PDF and the CDF can be expressed as \cite{Le}, \cite{Yac} 
\begin{align} \label{Ch2}
f_{X}(x)=\frac{\alpha\beta^{\alpha\mu}}{\Omega^{\alpha\mu}\Gamma(\mu)}x^{\alpha\mu-1}e^{-\left(\beta\frac{x}{\Omega}\right)^{\alpha}},
\end{align}
\noindent and
\begin{align} \label{Ch3}
F_{X}(x)=\frac{\gamma\left(\mu,\left(\beta\frac{x}{\Omega}\right)^{\alpha}\right)}{\Gamma(\mu)},
\end{align}
\noindent where $\beta \triangleq \frac{\Gamma\left(\mu+\frac{1}{\alpha}\right)}{\Gamma(\mu)}$, $\Omega=\mathbb{E}\{X\}$, and $\Gamma(\cdot)$ and $\gamma(\cdot,\cdot)$ denote the complete gamma function and the lower incomplete gamma function, respectively. Also, the $\nu^{th}$ moment of $X$ is given by
\begin{align} \label{Ch4}
\mathbb{E}\{|X|^{\nu}\}=\frac{\Gamma^{\nu-1}(\mu)\Gamma(\mu+\nu/\alpha)}{\Gamma^{\nu}(\mu+1/\alpha)}\Omega^{\nu}.
\end{align}
\noindent In this work, the $\alpha-\mu$ small-scale fading of the link from the STAR-RIS to the indoor user is modeled with parameters of $(\alpha_1,\mu_1,\Omega_1)$. Thus, the PDF and the CDF of $|\tilde{h}_{I,m}|$ can be expressed as (cf. [8]-[9]) 
\begin{align} \label{Ch40}
f_{|\tilde{h}_{I,m}|}(x)=\frac{\alpha_1\beta_1^{\alpha_1\mu_1}}{\Omega_1^{\alpha_1\mu_1}\Gamma(\mu_1)}x^{\alpha_1\mu_1-1}e^{-\left(\beta_1\frac{x}{\Omega_1}\right)^{\alpha_1}}, 
\end{align}
\noindent and
\begin{align} \label{Ch41}
F_{|\tilde{h}_{I,m}|}(x)=\frac{\gamma\left(\mu_1,\left(\beta_1\frac{x}{\Omega_1}\right)^{\alpha_1}\right)}{\Gamma(\mu_1)}.
\end{align}

\subsubsection{STAR-RIS to Outdoor User Channel Model}
For the outdoor environments, the small-scale fading at THz transmissions can be modeled by mixture of gamma or Gaussian mixture, as validated in \cite{Pap1}. When mixture of gamma is adopted, the PDF and CDF expression are given by
\begin{align} \label{Ch5}
f_{|\tilde{h}_{O,m}|}(x)=\sum_{n=1}^N a_{n}x^{b_n-1}e^{-c_nx},
\end{align}
\noindent and
\begin{align} \label{Ch6}
F_{|\tilde{h}_{O,m}|}(x)=\sum_{n=1}^N a_{n}c_n^{-b_n}\gamma(b_n,c_nx),
\end{align}
\noindent where $N$ is the number of mixture components, and $a_n, b_n,c_n$ are coefficients that satisfies $\sum_{n}^Na_{n}c_n^{-b_n}\Gamma(b_n)=1$ \cite{Le2}. Also, the $\nu^{th}$ moment of $|\tilde{h}_{O,m}|$ is given by
\begin{align} \label{Ch60}
\mathbb{E}\{|\tilde{h}_{O,m}|^{\nu}\}=\sum_{n=1}^N a_{n} \Gamma(b_n+\nu) c_n^{-(b_n+\nu)}.
\end{align} 
\noindent When Gaussian mixture is adopted, we have \cite{Sel}
\begin{align} \label{Ch7}
f_{|\tilde{h}_{O,m}|}(x)=\sum_{n=1}^N\varpi_{n}\frac{1}{\sqrt{2\pi}\eta_{n}}e^{-\frac{(x-\varkappa_{n})^2}{2\eta_{n}^2}},
\end{align}
\noindent and
\begin{align} \label{Ch8}
F_{|\tilde{h}_{O,m}|}(x)=\sum_{n=1}^N\varpi_{n}\left[1-Q\left(\frac{x-\varkappa_{n}}{\eta_{n}}\right)\right],
\end{align}
\noindent where $\varpi_{n}, \varkappa_{n}, \eta_{n}^2$ are the weight, the mean, and the variance of the $n^{th}$ component, respectively. Note that $\sum_{n=1}^N \varpi_{n}=1$, and $Q(\cdot)$ denotes a Gaussian Q-function. Also, the $\nu^{th}$ moment of $|\tilde{h}_{O,m}|$ in this case is given by \cite{Win}
\begin{align} \label{Ch80}
&\mathbb{E}\{|\tilde{h}_{O,m}|^{\nu}\}=\nonumber\\
&\left\{\begin{array}{lcr}
\sum\limits_{n=1}^N \varpi_{n}\eta_n^{\nu} 2^{\nu/2} \frac{\Gamma((\nu+1)/2)}{\sqrt{\pi}} {}_{1}F_1\left(-\frac{\nu}{2},\frac{1}{2},-\frac{\varkappa_n^2}{2\eta_n^2}\right),\\  \nu \mbox{\textit{ is even}} \\
\sum\limits_{n=1}^N \varpi_{n}\varkappa_n\eta_n^{\nu-1} 2^{(\nu+1)/2} \frac{\Gamma(\nu/2+1)}{\sqrt{\pi}} {}_{1}F_1\left(\frac{1-\nu}{2},\frac{3}{2},-\frac{\varkappa_n^2}{2\eta_n^2}\right), \\  \nu \mbox{\textit{ is odd}},\\
\end{array}\right.
\end{align}
\noindent where ${}_{1}F_1(\cdot)$ denotes the hypergeometric function. 

\subsubsection{AP to STAR-RIS Channel Model}
In a scenario of near-field where the indoor AP is located near the STAR-RIS\footnote{When the AP is located far from the STAR-RIS (i.e., a far-field scenario), a large number of RIS elements is deployed to compensate for the severe path-loss. In this case, gamma distribution can be used to characterize statistical distribution of the e2e channels from the AP to users, which is based on the Lyapunov central limit theorem and moment-matching for a distribution approximation. Further details can be found in our works of \cite[\textit{Appendix C}]{Le} and \cite[\textit{Appendix A}]{Le1}.}, the waves between the AP and the STAR-RIS can be modeled by spherical waves \cite{An3}. Specifically, the channel between the AP and the $m^{th}$ RIS element can be expressed as \cite{Li6}
\begin{align} \label{Ch4}
g_m=\sqrt{\Upsilon_m}e^{-j\frac{2\pi}{\lambda}d_m},
\end{align}
\noindent where $d_m=\sqrt{x_m^2+y_m^2+d_0^2}$ is the transmission distance from the AP to the $m^{th}$ RIS element, and $\Upsilon$ denotes the energy impinging on the $m^{th}$ element that can be calculated as\footnote{For a short transmission distance $d_m$, the value of the molecular absorption gain $e^{-\varrho(f)d_m/2}$ is very small, which is ignored for simplicity.} \cite{Li6}
\begin{align} \label{Ch40}
\Upsilon_m=\iint_{\mathcal{S}_m}\frac{G_t d_0^{\zeta+1}}{4\pi[x^2+y^2+d_0^2]^{(\zeta+3)/2}}dxdy,
\end{align} 
\noindent where $G_t=2(\zeta+1)$ is the antenna gain of the AP, and $\mathcal{S}_m=\{(x,y) \in \mathbb{R}^2\: x_m-\Delta_x/2 \leq x \leq x_m+\Delta_x/2, y_m-\Delta_y/2 \leq y \leq y_m+\Delta_y/2 \}$ with $\Delta_x$ and $\Delta_y$ denote the size of each RIS element along the x-axis and the y-axis directions, respectively. It is noted from Eq. (19) and Eq. (20) that the near-field modeling differs substantially from the far-field counterpart. Specifically, it is an element-wise distance-based model, where the angle, path length, and phase shift vary per RIS element. Also, RIS geometry knowledge (i.e., positions of RIS elements) are needed for the calculations of the transmission distance and the impinged energy. These have a profound impact on the channel estimation, RIS phase design, as well as beamforming design.

\subsection{Phase-Shift Adjustments}
In this work, we consider a coherent phase-shift design for both indoor and outdoor users. Recall that the phase shifts for reflection and transmission in the STAR-RIS with the ES protocol can be independently adjusted. For the reflecting side, the optimal phase-shifts of the STAR-RIS elements are designed to align the phases of the cascaded channels at the indoor user. Also, for the transmitting side, the optimal phase-shifts of the STAR-RIS elements are designed to align the phases of the cascaded channels at the outdoor user\footnote{When the same resource blocks are used for all users in a system with multiple indoor and multiple outdoor users, phase-shifts are designed for a specific goal, such as maximizing sum-rate or maximizing energy-efficiency, via alternative optimization, semidefinite relaxation, or machine learning. The readers are referred to [26]-[27], [29], and [31], for more details.}. Mathematically, the optimal phase-shifts of the $m^{th}$ element associated with the $\chi$ user are given by $\theta_{\chi,m}^{\star}=-\angle h_{\chi,m}-\angle g_{m}, \forall m \in \mathcal{M}, \chi=\{I,O\}$. Consequently, we can express the e2e channels as\footnote{We assume that perfect channel state information (CSI) is available in this work. Details about channel estimation in STAR-RIS-aided wireless systems can be found in \cite{Wu1}.} (cf. \eqref{S1})
\begin{align} \label{P1}
|H_{\chi}|&=\left|\sum_{m=1}^{M}h_{\chi,m}\mathrm{a}_{\chi,m}e^{j\theta_{\chi,m}^{\star}}g_{m}\right| \nonumber\\
&=\sum_{m=1}^M \bar{h}_{\chi,m}\bar{g}_{m}\mathrm{a}_{\chi,m}|\tilde{h}_{\chi,m}|=\sum_{m=1}^M \mathcal{A}_{\chi,m}|\tilde{h}_{\chi,m}|,
\end{align}  
\noindent where $\mathcal{A}_{\chi,m} \triangleq \bar{h}_{\chi,m}\bar{g}_{m}\mathrm{a}_{\chi,m}$. For the near-field scenario, we have $\bar{g}_m=\sqrt{\Upsilon}_m$ (cf. \eqref{Ch4}). It is worth noting that the coherent phase shifting design in \eqref{P1} guarantees the co-phasing of all arrived signals at the users. Moreover, it requires perfect phase alignment to achieve the desired signal combination.

\section{End-to-End (E2E) Channel Characterization}
In this section, we present results on a weighted sum of $\alpha-\mu$ variates, a weighted sum of mixture of gamma variates, and a weighted sum of Gaussian mixture variates for characterizing the e2e channels. These results are necessary for deriving the expressions of the performance metrics in the next section\footnote{The results would also be useful when analyzing the system from an physical layer security perspective, e.g., deriving secrecy outage probability and secrecy capacity}.

\subsection{Indoor User with $\alpha-\mu$ Fading Models}
The e2e channel of the indoor user is expressed as $|H_I|=\sum_{m=1}^M \mathcal{A}_{I,m}|\tilde{h}_{I,m}|$. Recall that $\tilde{h}_{I,m}|$ are independent and identically distributed (i.i.d.) $\alpha-\mu$ variates with parameters of $(\alpha_1, \mu_1, \Omega_1)$. Thus, $|H_I|$ is a weighted sum of $M$ i.i.d. $\alpha-\mu$ variates, or, equivalently, $|H_I|$ is a sum of $M$ independent but not identically distributed (i.n.i.d.) $\alpha-\mu$ variates with parameters of $(\alpha_1, \mu_1, \mathcal{A}_{I,m}\Omega_1)$.   

\noindent\textbf{\textit{Theorem 1:}} (\textbf{\textit{Approximate weighted sum of $\alpha-\mu$ variates}}) \textit{The PDF and CDF of a weighted sum of $\alpha-\mu$ variates can be approximated as}
\begin{align} \label{E1}
f_{|H_I|}(x)=\frac{\alpha_{\star}\beta_{\star}^{\alpha_{\star}\mu_{\star}}}{\Omega_{\star}^{\alpha_{\star}\mu_{\star}}\Gamma(\mu_{\star})}x^{\alpha_{\star}\mu_{\star}-1}e^{-\left(\beta_{\star}\frac{x}{\Omega_{\star}}\right)^{\alpha_{\star}}},
\end{align}
\noindent \textit{and}
\begin{align} \label{E2}
F_{|H_I|}(x)=\frac{\gamma\left(\mu_{\star},\left(\beta_{\star}\frac{x}{\Omega_{\star}}\right)^{\alpha_{\star}}\right)}{\Gamma(\mu_{\star})},
\end{align}
\noindent \textit{where $\alpha_{\star}$ and $\mu_{\star}$ are calculated via moment-based estimators.} 

\textit{Proof}: The results are obtained based on the proof in our work \cite[\textit{Appendix A}]{Le}.
   
In very recently, exact expressions of the PDF and CDF of a sum of i.i.d. $\alpha-\mu$ variates were derived in \cite{Gar}, which is based on theories of complex integration and differential equations. We now extend this framework to the case of i.n.i.d. distribution, which leads to the following result. 

\noindent\textbf{\textit{Theorem 2:}} (\textbf{\textit{Exact weighted sum of $\alpha-\mu$ variates}}) \textit{The PDF and CDF of a weighted sum of $\alpha-\mu$ variates are given by}
\begin{align} \label{E3}
f_{|H_I|}(x)=\left(\frac{\alpha_1\mu_1^{\mu_1}}{\Gamma(\mu_1)}\right)^M\sum_{i=0}^{\infty}\frac{\delta_ix^{i\alpha_1+M\alpha_1\mu_1-1}}{\Gamma(i\alpha_1+M\alpha_1\mu_1)},
\end{align}
\noindent \textit{and}
\begin{align} \label{E4}
F_{|H_I|}(x)=\left(\frac{\alpha_1\mu_1^{\mu_1}}{\Gamma(\mu_1)}\right)^M\sum_{i=0}^{\infty}\frac{\delta_ix^{i\alpha_1+M\alpha_1\mu_1}}{\Gamma(i\alpha_1+M\alpha_1\mu_1+1)},
\end{align}
\noindent \textit{where} 
\begin{align} \label{E41}
\delta_i=&\sum_{m_{M}=0}^i\mathcal{B}_{m_M,M}\sum_{m_{M-1}=0}^{i-m_M}\mathcal{B}_{m_{M-1},M-1}\cdots\nonumber\\
&\sum_{m_2=0}^{i-\sum_{j=3}^M{m_j}}\mathcal{B}_{m_2,2}\mathcal{B}_{i-\sum_{j=2}^M{m_j},1},
\end{align}
\noindent \textit{and} $\mathcal{B}_{n,m}=\frac{(-1)^n\beta_1^{\alpha_1(n+\mu_1)}\Gamma(\alpha_1(n+\mu_1))}{n!\mu_1^{\mu_1}(\mathcal{A}_{I,m}\Omega_1)^{\alpha_1(n+\mu_1)}}$.

\textit{Proof}: See Appendix A.

\begin{table}[!t]
\scriptsize
\caption{Truncation Error.}
\vspace{10pt}
\centering
\begin{tabular}{|@{\hskip3pt}l@{\hskip3pt}|@{\hskip3pt}l@{\hskip3pt}|@{\hskip3pt}l@{\hskip3pt}|}
\hline
\hfil \textbf{Scenarios} & \hfil \textbf{Truncation error} & \hfil \textbf{Truncation error} \\
\hfil & \hfil \textbf{of the PDF} & \hfil \textbf{of the CDF} \\
\hline
M = 2: $\mathcal{A}_I=[1, 0.7]$  & $1.5786 \times 10^{-17}$ & $1.9189 \times 10^{-18}$ \\
\hline
M = 3: $\mathcal{A}_I=[1, 0.7, 2.5]$  & $6.3065 \times 10^{-17}$ & $7.3328 \times 10^{-18}$ \\
\hline
M = 4: $\mathcal{A}_I=[1, 0.7, 2.5, 1.4]$  & $1.1035 \times 10^{-15}$ & $1.2298 \times 10^{-16}$ \\
\hline
M = 5: $\mathcal{A}_I=[1, 0.7, 2.5, 1.4, 0.8]$  & $8.2042 \times 10^{-14}$ & $8.7799 \times 10^{-15}$ \\
\hline
\end{tabular}
\end{table}

\begin{figure}[!t] 
	\centering{\includegraphics[width=0.45\textwidth]{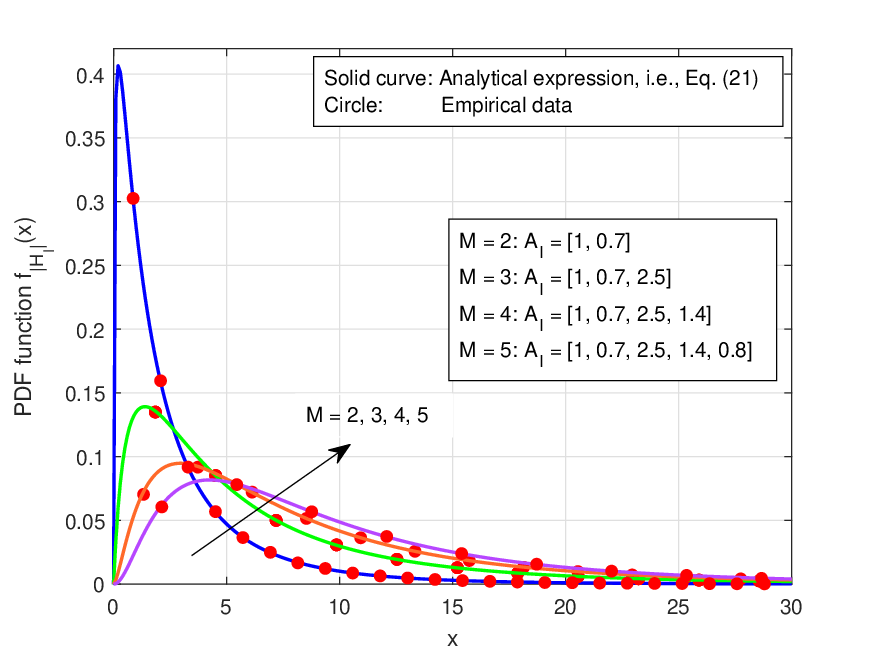}}
	\caption{PDF of $|H_I|$ under different values of $M$.}
	\label{Fig2}
\end{figure}

\begin{figure}[!t] 
	\centering{\includegraphics[width=0.45\textwidth]{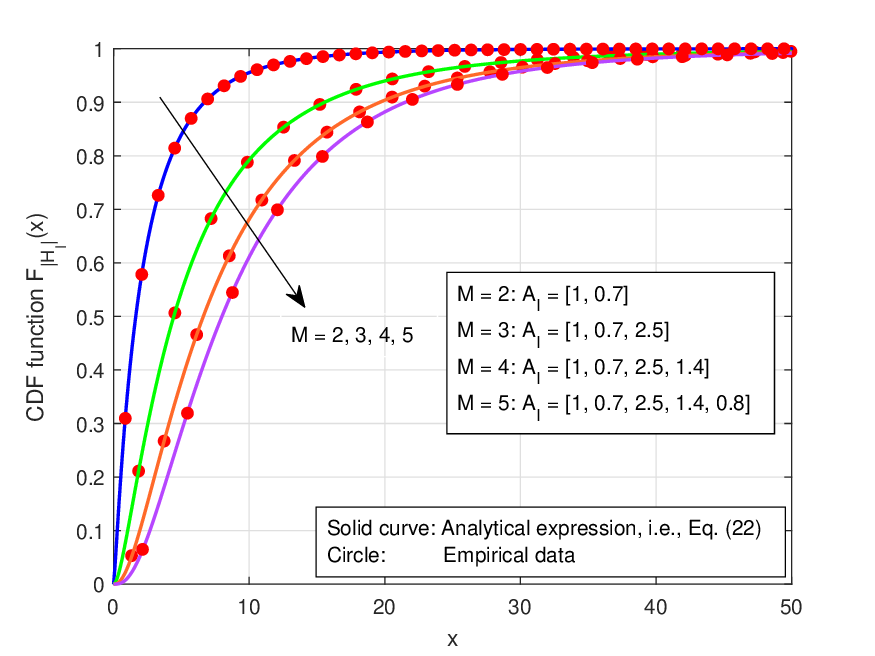}}
	\caption{CDF of $|H_I|$ under different values of $M$.}
	\label{Fig3}
\end{figure}

\noindent As shown in \textit{Appendix A}, $\delta_i$ can be calculated in a recursive manner. Then, \eqref{E3} and \eqref{E4} are easily obtained. In Fig. 2 and Fig. 3, we plot the PDF and the CDF curves to demonstrate the accuracy of the above expressions. Here, we use the same $\alpha-\mu$ parameters as in \cite{Gar}, i.e., $\alpha_1=0.5, \mu_1=1.5$, and $\hat{x}=1$. The results confirm the high accuracy of the derived expressions of \eqref{E3} and \eqref{E4}. It is also worth mentioning that while the PDF and CDF formulas above are expressed in infinite forms, high accuracy can be achieved with a quite small number of terms. Specifically, we show in Table 1 the truncation errors of the PDF and the CDF at $x=2$ using $\mathcal{N}_T =30$ terms. It can be seen that no more than 30 terms are required to guarantee a remarkable accuracy of less than $10^{-13}$ in all scenarios.

\subsection{Outdoor User with Mixture Fading Models}
	
For the outdoor user, we have $|H_O|=\sum_{m=1}^M \mathcal{A}_{O,m}|\tilde{h}_{O,m}|$, where $\mathcal{A}_{O,m} \triangleq \bar{h}_{O,m}\bar{g}_{m}\mathrm{a}_{O,m}$. When a mixture of gamma model is used for $|\tilde{h}_{O,m}|$, we have the result below.

\noindent\textbf{\textit{Theorem 3:}} \cite{Le1} (\textbf{\textit{Approximate weighted sum of mixture of gamma variates}}) \textit{The PDF and CDF of a weighted sum of mixture of gamma variates can be approximated as}
\begin{align} \label{E5}
f_{|H_O|}(x)=\widetilde{\sum_{\mathbf{n}}}\tilde{a}_{\mathbf{n}}x^{\tilde{b}_{\mathbf{n}}-1}e^{-\tilde{c}_{\mathbf{n}}x},
\end{align}
\noindent \textit{and}
\begin{align} \label{E6}
F_{|H_O|}(x)=\widetilde{\sum_{\mathbf{n}}}\tilde{a}_{\mathbf{n}}\tilde{c}_{\mathbf{n}}^{-\tilde{b}_{\mathbf{n}}}\gamma(\tilde{b}_{\mathbf{n}},\tilde{c}_{\mathbf{n}}x),
\end{align}
\noindent \textit{where $\widetilde{\sum_{\mathbf{n}}} \triangleq \sum_{n_1=1}^N\sum_{n_2=1}^N\cdots\sum_{n_M=1}^N$, $\tilde{a}_{\mathbf{n}}= \frac{\Phi_{3,\mathbf{n}}[\tilde{c}_{\mathbf{n}}]^{\tilde{b}_{\mathbf{n}}}}{\Gamma(\tilde{b}_{\mathbf{n}})}$, $\tilde{b}_{\mathbf{n}}=\frac{\Phi_{1,\mathbf{n}}^2}{\Phi_{2,\mathbf{n}}}$, $\tilde{c}_{\mathbf{n}}=\frac{\Phi_{1,\mathbf{n}}}{\Phi_{2,\mathbf{n}}}$, $\Phi_{1,\mathbf{n}}=\sum_{m=1}^M b_{n_m}\frac{\mathcal{A}_{O,m}}{c_{n_m}}$, $\Phi_{2,\mathbf{n}}=\sum_{m=1}^M b_{n_m}\left(\frac{\mathcal{A}_{O,m}}{c_{n_m}}\right)^2$, and $\Phi_{3,\mathbf{n}}=\prod_{m=1}^M\frac{a_{n_m}\Gamma(b_{n_m})}{c_{n_m}^{b_{n_m}}}$}.

On the other hand, if a Gaussian mixture is adopted for modeling $|\tilde{h}_{O,m}|$, we have the following result.
   
\noindent\textbf{\textit{Theorem 4:}} (\textbf{\textit{Approximate weighted sum of Gaussian mixture variates}}) \textit{The PDF and CDF of a weighted sum of Gaussian mixture variates can be approximated as}
\begin{align} \label{E7}
f_{|H_O|}(x)=\widetilde{\sum_{\mathbf{n}}}\tilde{\varpi}_{\mathbf{n}}\frac{1}{\sqrt{2\pi}\tilde{\eta}_{\mathbf{n}}}e^{-\frac{(x-\tilde{\varkappa}_{\mathbf{n}})^2}{2\tilde{\eta}_{\mathbf{n}}^2}},
\end{align}
\noindent and
\begin{align} \label{E8}
F_{|H_O|}(x)=\widetilde{\sum_{\mathbf{n}}}\tilde{\varpi}_{\mathbf{n}}\left[1-Q\left( \frac{x-\tilde{\varkappa}_{\mathbf{n}}}{\tilde{\eta}_{\mathbf{n}}}\right)\right],
\end{align}
\noindent \textit{where $\tilde{\varpi}_{\mathbf{n}} \triangleq \prod_{m=1}^M\varpi_{n_m}$, $\tilde{\varkappa}_{\mathbf{n}} \triangleq \sum_{m=1}^M\mathcal{A}_{O,m}\varkappa_{n_m}$, and $\tilde{\eta}_{\mathbf{n}}^2 \triangleq \sum_{m=1}^M\mathcal{A}_{O,m}^2\eta_{n_m}^2$}. 

\textit{Proof}: See Appendix B.

\noindent The accuracy of the expressions of \eqref{E7} and \eqref{E8} are illustrated in Fig. 4. Note that the parameters of Gaussian mixture variates (i.e., the weight, the mean and the variance of each component) used to obtain this result are based on \cite[Table III]{Sel}.

\begin{figure}[t] 
	\centering{\includegraphics[width=0.45\textwidth]{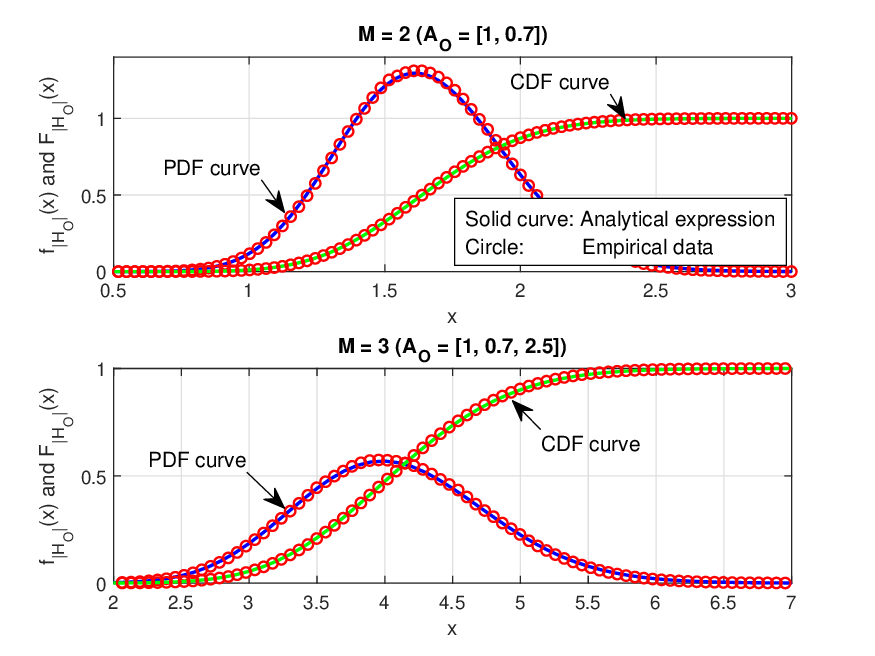}}
	\caption{PDF and CDF of $|H_O|$.}
	\label{Fig4}
\end{figure}

\section{Analysis of Outage Probability and Capacity}
\subsection{Analysis of Outage Probability}
The outage probability (OP) is conventionally defined as
\begin{align} \label{O1}
P_{out}=Pr(\gamma<\gamma_{th}),
\end{align} 
\noindent where $\gamma$ is the received SNR, and $\gamma_{th}$ denotes the SNR threshold. With the help of the statistical distributions of the e2e derived in Section III, we obtain the closed-form expressions of the OPs of the users.

\noindent\textbf{\textit{Theorem 5:}} \textit{The OP of the indoor user is given by}
\begin{align} \label{O2}
P_{out,I}(\gamma_{th,I})=\left\{\begin{array}{lcl}
\left(\frac{\alpha_1\mu_1^{\mu_1}}{\Gamma(\mu_1)}\right)^M\sum_{i=0}^{\infty}\frac{\delta_i[\Psi_I(\gamma_{th,I})]^{(i\alpha_1+M\alpha_1\mu_1)/2}}{\Gamma(i\alpha_1+M\alpha_1\mu_1+1)}, \\ \quad \mbox{     \textit{if}} \quad \gamma_{th,I} < \frac{\rho_I}{\kappa^2} \mbox{     \textit{and} } \gamma_{th,O} < \frac{\rho_O}{\kappa^2+\rho_I}\\
1, \quad \mbox{\textit{if}} \quad \gamma_{th,I} \geq \frac{\rho_I}{\kappa^2} \mbox{     \textit{or} } \gamma_{th,O} \geq \frac{\rho_O}{\kappa^2+\rho_I},\\
\end{array}\right.
\end{align}
\noindent \textit{where} $\Psi_I(\gamma_{th,I}) \triangleq \max\left\{\Psi_{IO}(\gamma_{th,I}), \Psi_O(\gamma_{th,O}) \right\}$, $\Psi_{IO}(\gamma_{th,I}) \triangleq \frac{N_0\gamma_{th,I}}{P_I-\gamma_{th,I}\kappa^2 P}$, $\Psi_O(\gamma_{th,O}) \triangleq \frac{N_0\gamma_{th,O}}{P_O-\gamma_{th,O}(\kappa^2 P+P_I)}$, $\gamma_{th,I}$ \textit{and} $\gamma_{th,O}$ \textit{are the SNR thresholds of the indoor and outdoor users, respectively}, \textit{and} $\delta_i$ \textit{is a coefficient defined as in \eqref{E41}. Also, the OP of the outdoor user is given by}
\begin{align} \label{O3}
P_{out,O}(\gamma_{th,O})=\left\{\begin{array}{lcl}
\widetilde{\sum\limits_{\mathbf{n}}}\tilde{a}_{\mathbf{n}}\tilde{c}_{\mathbf{n}}^{-\tilde{b}_{\mathbf{n}}}\gamma(\tilde{b}_{\mathbf{n}},\tilde{c}_{\mathbf{n}}\sqrt{\Psi_O(\gamma_{th,O})}), \\
 \quad \mbox{      \textit{if} } \gamma_{th,O} < \frac{\rho_O}{\kappa^2+\rho_I} \\
1, \quad \mbox{\textit{if} } \gamma_{th,O} \geq \frac{\rho_O}{\kappa^2+\rho_I},\\
\end{array}\right.
\end{align}
\noindent \textit{for the mixture of gamma case, and}
\begin{align} \label{O30}
P_{out,O}(\gamma_{th,O})=\left\{\begin{array}{lcl}
\widetilde{\sum\limits_{\mathbf{n}}}\tilde{\varpi}_{\mathbf{n}}\left[1-Q\left( \frac{\sqrt{\Psi_O(\gamma_{th,O})}-\tilde{\varkappa}_{\mathbf{n}}}{\tilde{\eta}_{\mathbf{n}}}\right)\right], \\
\quad \mbox{      \textit{if} } \gamma_{th,O} < \frac{\rho_O}{\kappa^2+\rho_I} \\
1, \quad \mbox{\textit{if} } \gamma_{th,O} \geq \frac{\rho_O}{\kappa^2+\rho_I},\\
\end{array}\right.
\end{align}
\noindent \textit{for the Gaussian mixture case.}

\textit{Proof}: For the indoor user with SIC, the outage occurs when it cannot decode the signal of the outdoor user or its own signal. Thus, the OP is defined as $P_{out,I}(\gamma_{th,I})=1-Pr(\gamma_{I}>\gamma_{th,I}, \gamma_{I \rightarrow O}>\gamma_{th,O})=1-Pr\left(\frac{P_I|H_I|^2}{\kappa^2 P|H_I|^2+N_0}>\gamma_{th,I}, \frac{P_O|H_I|^2}{(\kappa^2 P+P_I)|H_I|^2+N_0}>\gamma_{th,O}\right)$. If $\gamma_{th,I} < \frac{\rho_I}{\kappa^2}$ and $\gamma_{th,O} < \frac{\rho_O}{\kappa^2+\rho_I}$, we arrive at $P_{out,I}(\gamma_{th,I})=F_{|H_I|}\left(\sqrt{\Psi_I(\gamma_{th,I})}\right)$. Otherwise, we have $P_{out,I}(\gamma_{th,I})=1$. With the help of \eqref{E4}, we obtain \eqref{O2}. For the case of outdoor user, it is straightforward to show that $P_{out,O}(\gamma_{th,O})=F_{|H_O|}\left(\sqrt{\Psi_O(\gamma_{th,O})}\right)$ if $\gamma_{th,O} < \frac{\rho_O}{\kappa^2+\rho_I}$. By using \eqref{E6} and \eqref{E8}, we obtain \eqref{O3} and \eqref{O30}. This completes the proof.  \hspace*{\fill} $\blacksquare$

\textit{\textbf{Asymptotic at high SNR}:} At the high SNR regime (i.e., $\bar{\gamma} \triangleq P/N_0 \rightarrow \infty$), we have $\Psi_I(\gamma_{th,I})=\max\left\{\frac{\gamma_{th,I}}{\bar{\gamma}[\rho_I-\gamma_{th,I}\kappa^2]},\frac{\gamma_{th,O}}{\bar{\gamma}[\rho_O-\gamma_{th,O}(\kappa^2 +\rho_I)]}\right\} \rightarrow 0$ and $\Psi_O(\gamma_{th,O})=\frac{\gamma_{th,O}}{\bar{\gamma}[\rho_O-\gamma_{th,O}(\kappa^2 +\rho_I)]} \rightarrow 0$. Thus, by taking the first term of the summation in \eqref{O2}, we can express the asymptotic OP of the indoor user in the near-field scenario as
\begin{align} \label{O5}
P_{out,I}^{\infty}(\gamma_{th,I})\! =\! \left(\!\frac{\alpha_1\mu_1^{\mu_1}}{\Gamma(\mu_1)}\!\right)^{\!M}\!\frac{\delta_0[\Psi_I(\gamma_{th,I})]^{\frac{M\alpha_1\mu_1}{2}}}{\Gamma(M\alpha_1\mu_1+1)}\propto \bar{\gamma}^{-\frac{M\alpha_1\mu_1}{2}}.
\end{align}

\noindent For the outdoor user with the mixture of gamma distribution, we can adopt an approximation of $\gamma(c,x) \overset{x \rightarrow 0}{\approx} x^c/c$ for asymptotic expression. In particular, we have 
\begin{align} \label{O6}
P_{out,O}^{\infty}(\gamma_{th,O}) = \widetilde{\sum\limits_{\mathbf{n}}} \frac{\tilde{a}_{\mathbf{n}}}{\tilde{b}_{\mathbf{n}}} [\Psi_O(\gamma_{th,O})]^{\tilde{b}_{\mathbf{n}}/2}.
\end{align}

\subsection{Analysis of Ergodic Capacity}
The ergodic capacity (in \textit{bit/s/Hz}) can be evaluated by
\begin{align} \label{C1}
C_e=\mathbb{E}\{\log_2(1+\gamma)\}=\int_0^{\infty}\log_2(1+x)f_{\gamma}(x)dx,
\end{align}
\noindent where $f_{\gamma}(x)$ is the PDF of $\gamma$. With the statistical distributions obtained in Section III, we can evaluate \eqref{C1} to obtain the following results.

\noindent\textbf{\textit{Theorem 6:}} \textit{The ergodic capacity expression of the indoor user is given by}
\begin{align} \label{C2}
C_{e,I}\approx \frac{1}{\Gamma(\mu_{\star})}\sum_{q=1}^Q \! w_q t_q^{\mu_{\star}-1}\log_2\left(\!1\!+\!\frac{t_q^{2/\alpha_{\star}}\Omega_{\star}^2P_I}{N_0\beta_{\star}^2\!+\!t_q^{2/\alpha_{\star}}\Omega_{\star}^2\kappa^2P}\!\right)\!,
\end{align}
\noindent \textit{where} $w_q$ \textit{and} $t_q$ ($q=1,2,...,Q$) \textit{are the weights and abscissas of the $Q$-point Gauss-Laguerre quadrature. Also, the capacity of the outdoor user with the mixture of gamma model is} 
\begin{align} \label{C3}
&C_{e,O} \approx \nonumber\\
&\!\widetilde{\sum_{\mathbf{n}}}\tilde{a}_{\mathbf{n}}\tilde{c}_{\mathbf{n}}^{-\tilde{b}_{\mathbf{n}}}\!\sum_{q=1}^Q\! w_q t_q^{\tilde{b}_{\mathbf{n}}-1}\!\log_2\!\!\left(\!\!1\!+\!\frac{t_q^2P_O}{N_0\tilde{c}_{\mathbf{n}}^2\!+\!t_q^2(\kappa^2P\!+\!P_I)}\!\right)\!.
\end{align}
\noindent \textit{For the Gaussian mixture model, the capacity is given by}
\begin{align} \label{C30}
C_{e,O} \approx &\widetilde{\sum\limits_{\mathbf{n}}}\tilde{\varpi}_{\mathbf{n}}\frac{1}{\sqrt{\pi}}e^{-\frac{\tilde{\varkappa}_{\mathbf{n}}^2}{2\tilde{\eta}_{\mathbf{n}}^2}}\sum\limits_{q=1}^Q w_q e^{-t_q^2+t_q+\frac{\tilde{\varkappa}_{\mathbf{n}}\sqrt{2}}{\tilde{\eta}_{\mathbf{n}}}t_q}\nonumber\\
&\times \log_2\left(1+\frac{2\tilde{\eta}_{\mathbf{n}}^2t_q^2P_O}{N_0+2\tilde{\eta}_{\mathbf{n}}^2t_q^2(\kappa^2P+P_I)}\right).
\end{align}

\textit{Proof}: See Appendix C.

\begin{table}[!t]
\scriptsize
\begin{center}
\captionsetup{font=scriptsize}
\caption{Simulation parameters.}
\vspace{10pt}
\begin{tabular}{|@{\hskip3pt}l@{\hskip3pt}|@{\hskip3pt}l@{\hskip3pt}|}
\hline
\hfil \textbf{Parameters} & \hfil \textbf{Values}\\
\hline
Operating frequency  & $f = 140$ (GHz) \cite{Pap}, \cite{Pap1} \\
Transmission bandwidth  & $4$ (GHz) \cite{Pap}, \cite{Pap1} \\
Molecular absorption coefficient & $\varrho = 3.18\times10^{-4}$ per meter \cite{Le}, \cite{Cha}\\
Hardware impairment level & $\kappa^2=0.08$ \cite{Le}, \cite{Le1} \\
Antenna gains & $G_t=20$, $G_r=20$ (dBi) \cite{Le1} \\
Power allocation & $\rho_I=0.4$, $\rho_O=0.6$ \cite{Zha1}\\
STAR-RIS panel & $M = 9$ elements \\
Sizes of each RIS element & $\Delta_x = \Delta_y =1$ (cm) \\
SNR threshold & $\gamma_{th,I}=\gamma_{th,O}=0$ (dB)\\
Power splitting coefficients & $\mathrm{a}_{I,m}^2=\mathrm{a}_{O,m}^2=0.5$ \cite{Li6}\\ 
Noise PSD & $-174$ (dBm/Hz) \\
\hline
\end{tabular}
\end{center}
\end{table}

\textit{\textbf{Asymptotic at high SNR}:} At the high SNR regime, we can simplify the capacity expressions as (see \textit{Appendix C} for details) 
\begin{align} \label{C4}
C_{e,I}^{\infty}=\left\{\begin{array}{lcl}
\log_2\left(1+\frac{\rho_I}{\kappa^2}\right) &, \kappa^2\neq 0 \\
\frac{2\psi(\mu_{\star})}{\alpha_{\star}\ln 2}+\log_2\left(\frac{\Omega_{\star}^2P_I}{\beta_{\star}^2N_0}\right) &, \kappa^2 = 0.\\
\end{array}\right.
\end{align} 
\noindent where $\psi(\cdot)$ is the psi function \cite[Eq. (8.360.1)]{Gra}, and
\begin{align} \label{C5}
C_{e,O}^{\infty}=\log_2\left(\frac{1+\kappa^2}{\rho_I+\kappa^2}\right).
\end{align}

\textit{\textbf{Asymptotic at low SNR}:} For THz communication systems, low SNR regime would be of interest and prevalent in realistic scenarios. To this end, we perform mathematical analysis of the achieved capacity in the low SNR regime in this subsection. Specifically, we adopt the second-order Taylor approximation to derive the asymptotic expression of the capacity \cite{Ver}, which leads to the following result. 

\noindent\textbf{\textit{Theorem 7:}} \textit{The asymptotic expression of the ergodic capacities at the low SNR regime is given by}
\begin{align} \label{D20}
C_{e,\chi}^{0} \approx &\left(\log_2e\right)\left[\left(\frac{PG_{\chi}}{N_0}\right)\rho_{\chi}\mathbb{E}\left\{|\tilde{H}_{\chi}|^2\right\}-\right. \nonumber\\
&\left.\left(\frac{PG_{\chi}}{N_0}\right)^2\rho_{\chi}(2\kappa^2+\rho_I+\varrho)\mathbb{E}\left\{|\tilde{H}_{\chi}|^4\right\}\right],
\end{align}      
\noindent \textit{where $\varrho = 0$ for the indoor user (i.e., $\chi=I$) and $\varrho = 1$ for the outdoor user (i.e., $\chi=O$), and $G_{\chi} \triangleq \min\{\mathcal{A}_{\chi,m}^2\}, \forall m$. Also, the values of $\mathbb{E}\left\{|\tilde{H}_{\chi}|^2\right\}$ and $\mathbb{E}\left\{|\tilde{H}_{\chi}|^4\right\}$ are given in Appendix D.} 

\textit{Proof}: See Appendix D.

\noindent \textit{\textbf{Remarks}}:   

\noindent 1) The results of \textit{Theorem 5} show that a higher hardware impairment level $\kappa^2$ leads to an increased outage probability for both indoor and outdoor users. 
 
\noindent 2) For the ergodic capacities, in the presence of HWI (i.e., $\kappa^2\neq 0$), the capacities of both users tend to be saturated at values determined by the HWI level $\kappa^2$ and the power allocation coefficients (i.e., $\rho_I$ and $\rho_O$). On the other hand, in case of no HWI (i.e., $\kappa^2 = 0$), the capacity of the indoor user $C_{e,I}^{\infty}$ increases logarithmically with the SNR. Meanwhile, the capacity of the outdoor user is governed by the power allocation coefficient $\rho_I$ due to the power-domain NOMA principle.

\noindent 3) Asymptotic analysis at the high SNR reveals that the diversity order of the OP of the indoor user is $d = -M\alpha_1\mu_1/2$ , which is not affected by the HWI level $\kappa^2$ (cf. Eq. (35)). Meanwhile, for the ergodic capacity, the high SNR slope in the ideal case (i.e., $\kappa^2 = 0$) is one, whilst the slope when $\kappa^2 \neq 0$ becomes zero since the capacity is saturated in the presence of HWI (cf. Eq. (41)). For the outdoor user, the high SNR slope of the capacity is zero since $C_{e,O}^{\infty}$ is a constant (cf. Eq. (42)). 

\noindent 4) For the STAR-RIS OMA system, the OP expressions of the indoor user and the outdoor user, denoted by $P_{out,\chi}^{OMA}(\gamma_{th,\chi}^{OMA})$, are obtained based \textit{Theorem 5}, where $\Psi_I(\gamma_{th,I})$ and $\Psi_O(\gamma_{th,O})$ in Eq. (32) - Eq. (34) are replaced by $\Psi_I^{OMA}(\gamma_{th,I}^{OMA}) \triangleq \frac{N_0\gamma_{th,I}^{OMA}}{P_I-\gamma_{th,I}^{OMA}\kappa^2 P}$ and $\Psi_O^{OMA}(\gamma_{th,O}^{OMA}) \triangleq \frac{N_0\gamma_{th,O}^{OMA}}{P_O-\gamma_{th,O}^{OMA}\kappa^2 P}$, respectively. Also, the condition that $P_{out,\chi}^{OMA}(\gamma_{th,\chi}^{OMA})=1$ becomes $\gamma_{th,\chi}^{OMA} \geq \frac{\rho_{\chi}}{\kappa^2}$. Similarly, the ergodic capacity expressions in the STAR-RIS OMA are also based \textit{Theorem 6}, where the term $(\kappa^2 P+P_I)$ in Eq. (39) - Eq. (40) is replaced by $\kappa^2 P$. Note that the capacity values need to be divided by 2 given that $C_{e,\chi}^{OMA} \triangleq \frac{1}{2}\mathbb{E}\{\log_2(1+\gamma_{\chi}^{OMA})\}$.

\section{Simulation Results and Discussions} 
In this section, we provide simulation results for the performance evaluation. The simulation parameters are listed in Table II. The coordinates of the STAR-RIS, the AP, the indoor user, and the outdoor user are $(0,0,0)$, $(0,0,-1)$, $(5,-2,-9)$ and $(7,-3,15)$, respectively. Unless otherwise specified, for the indoor environment, the small-scale fading coefficients of the RIS-to-Indoor User link is $\alpha_{1}=2$, $\mu_{1}=1$, $\Omega_{1}=1$ \cite{Le}, \cite{Cha}. For the outdoor environment, the small-scale fading of mixture of gamma is adopted for demonstrations\footnote{The coefficients of the mixture of gamma, i.e., $\{a_n, b_n, c_n\}$, is obtained based on the Rician fading channel via a method developed in \cite{Ata}. Also, the Rician fading channel model comprising of both LoS and non-LoS paths with the Rician factor of $K = 1$ is adopted for the outdoor environment, which is similar to \cite{Hua}.}. Also, the ES protocol is adopted for the STAR-RIS panel.

\begin{figure}[t] 
	\centering{\includegraphics[width=0.45\textwidth]{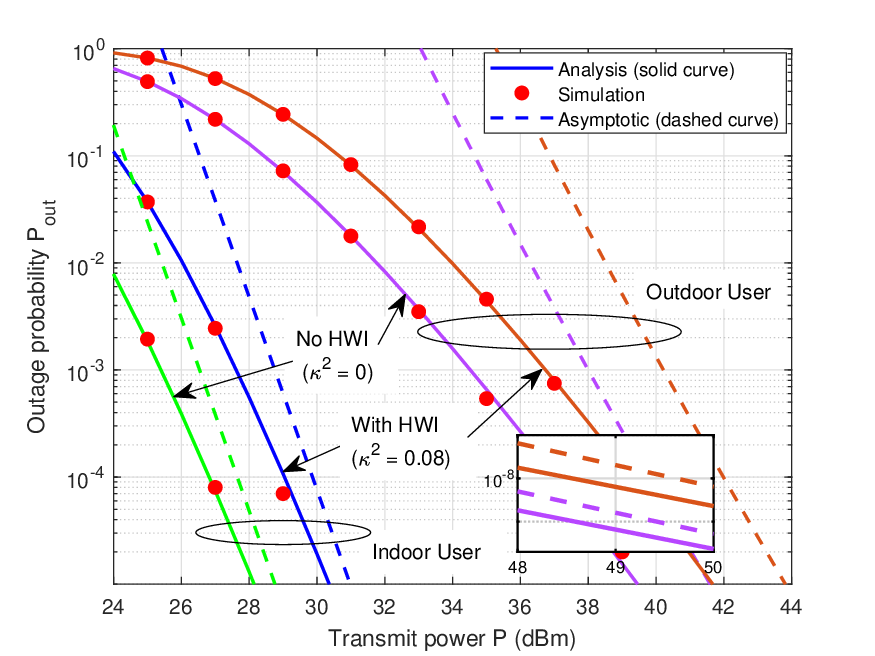}}
	\caption{Outage probabilities versus transmit power.}
	\label{Fig5}
\end{figure}

\begin{figure}[t] 
	\centering{\includegraphics[width=0.45\textwidth]{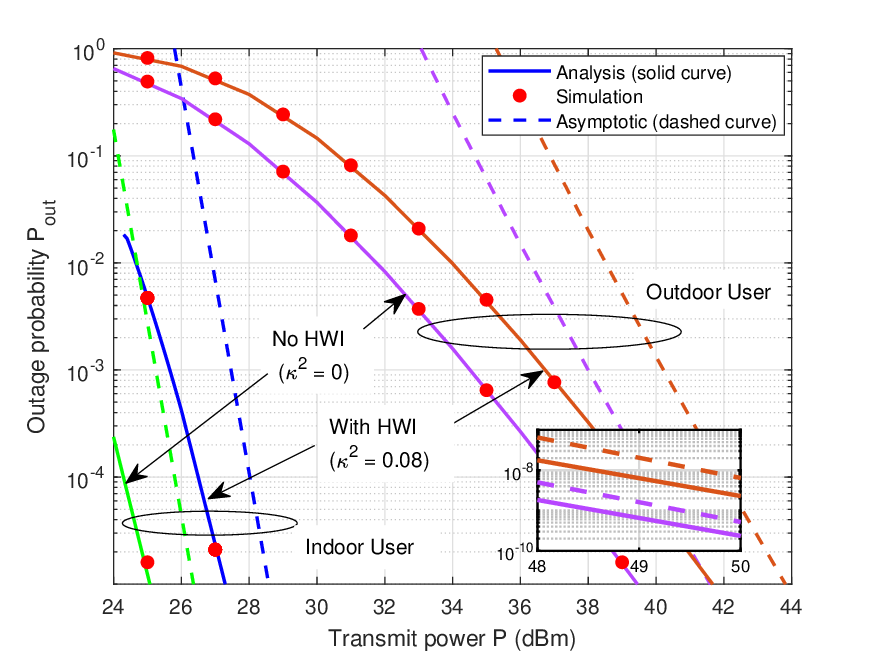}}
	\caption{Outage probability under a different fading condition.}
	\label{Fig6}
\end{figure}

\begin{figure}[t] 
	\centering{\includegraphics[width=0.45\textwidth]{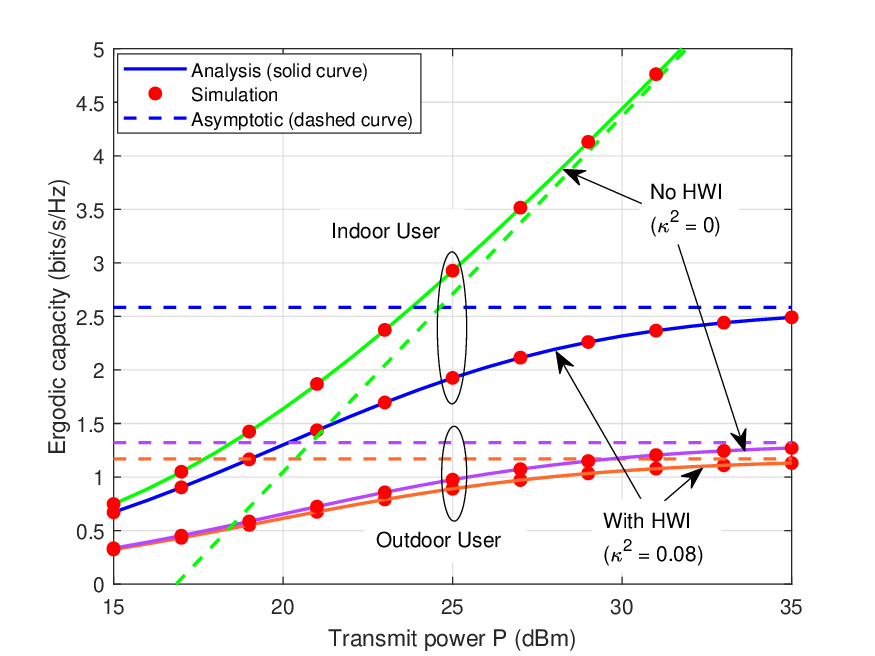}}
	\caption{Ergodic capacities versus transmit power.}
	\label{Fig7}
\end{figure}

\begin{figure}[t] 
	\centering{\includegraphics[width=0.45\textwidth]{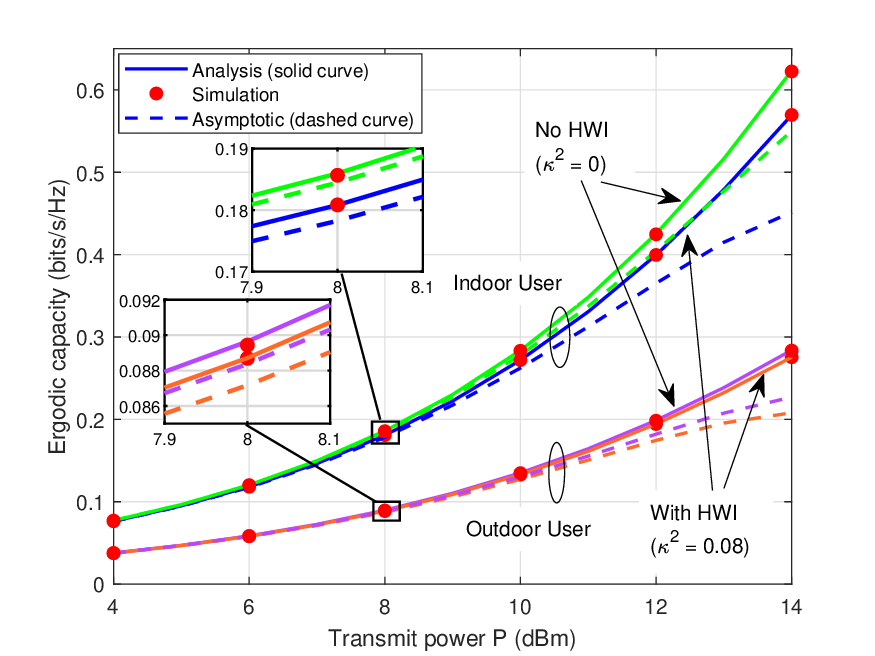}}
	\caption{Ergodic capacities at low SNR regime.}
	\label{Fig8}
\end{figure}

\begin{figure}[t] 
	\centering{\includegraphics[width=0.45\textwidth]{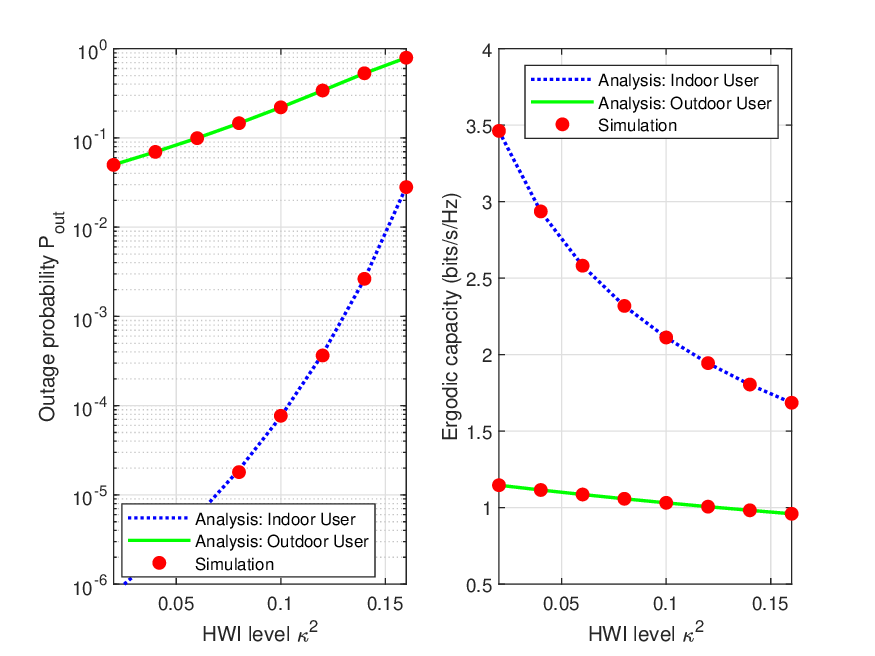}}
	\caption{OP and capacities versus HWI level ($P = 30$ (dBm)).}
	\label{Fig9}
\end{figure}

\begin{figure}[t] 
	\centering{\includegraphics[width=0.45\textwidth]{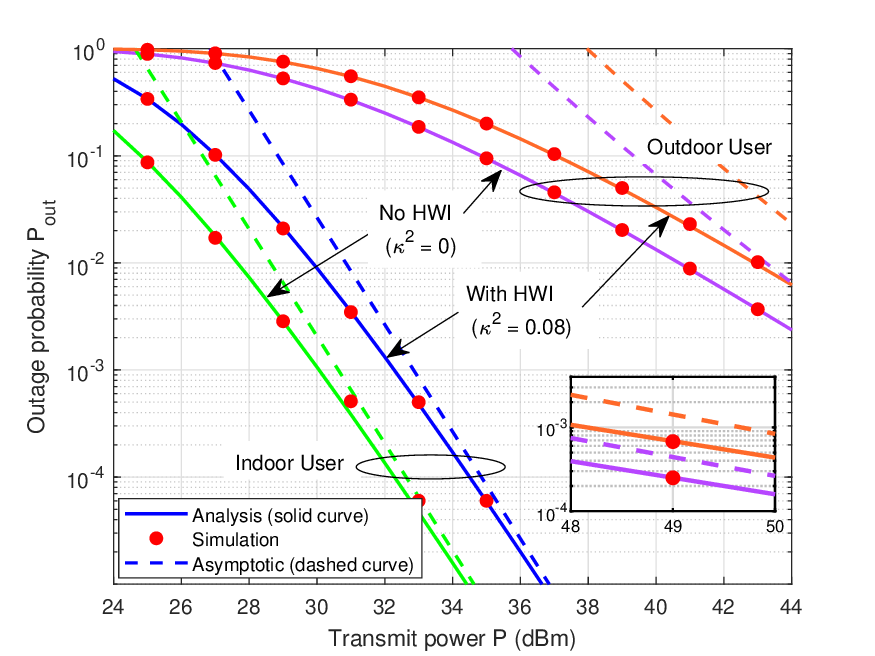}}
	\caption{Outage probabilities in MS mode.}
	\label{Fig10}
\end{figure}

\begin{figure}[t] 
	\centering{\includegraphics[width=0.45\textwidth]{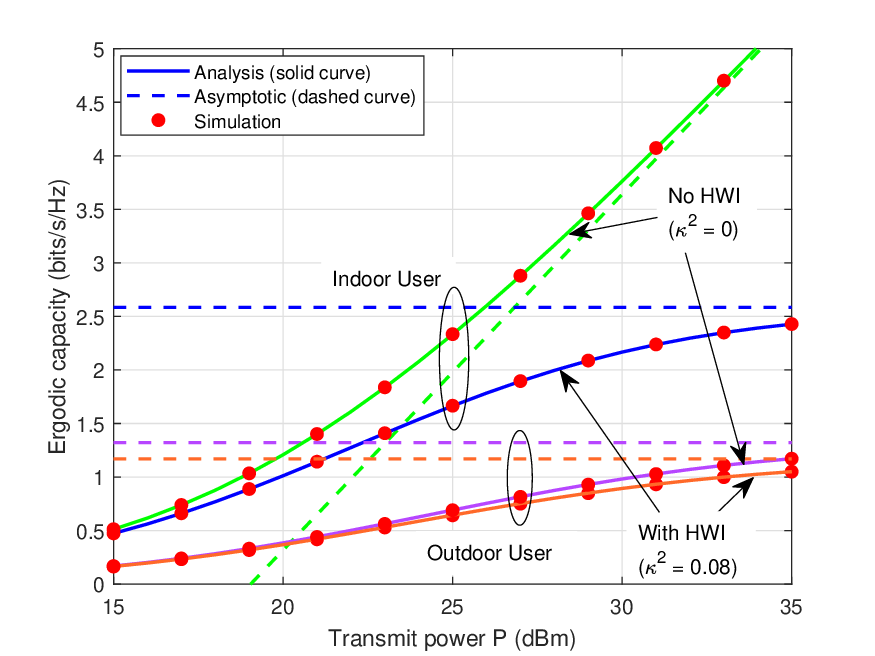}}
	\caption{Ergodic capacities in MS mode.}
	\label{Fig11}
\end{figure}

\begin{figure}[t] 
	\centering{\includegraphics[width=0.45\textwidth]{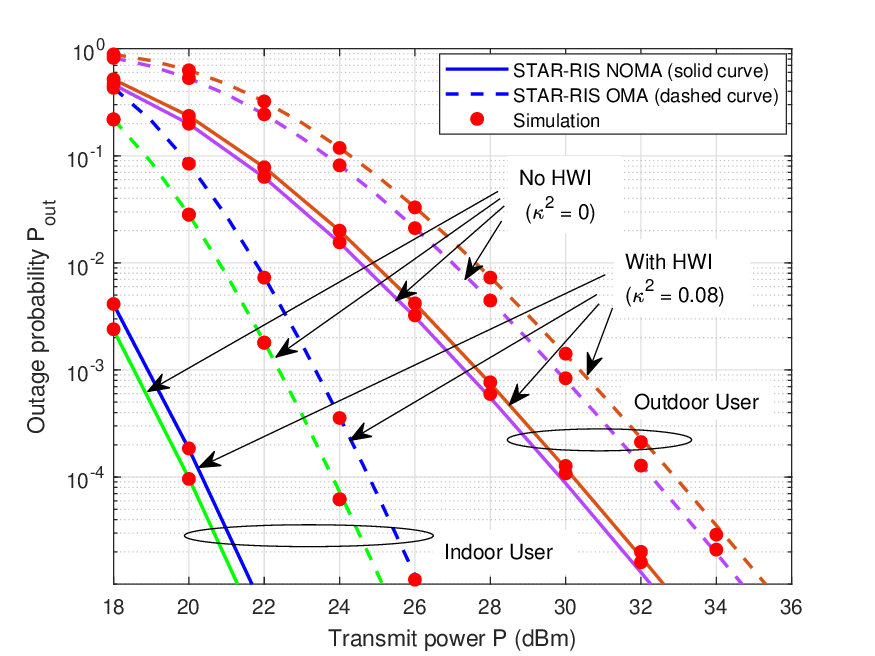}}
	\caption{Outage probabilities: NOMA versus OMA.}
	\label{Fig12}
\end{figure}

\begin{figure}[t] 
	\centering{\includegraphics[width=0.45\textwidth]{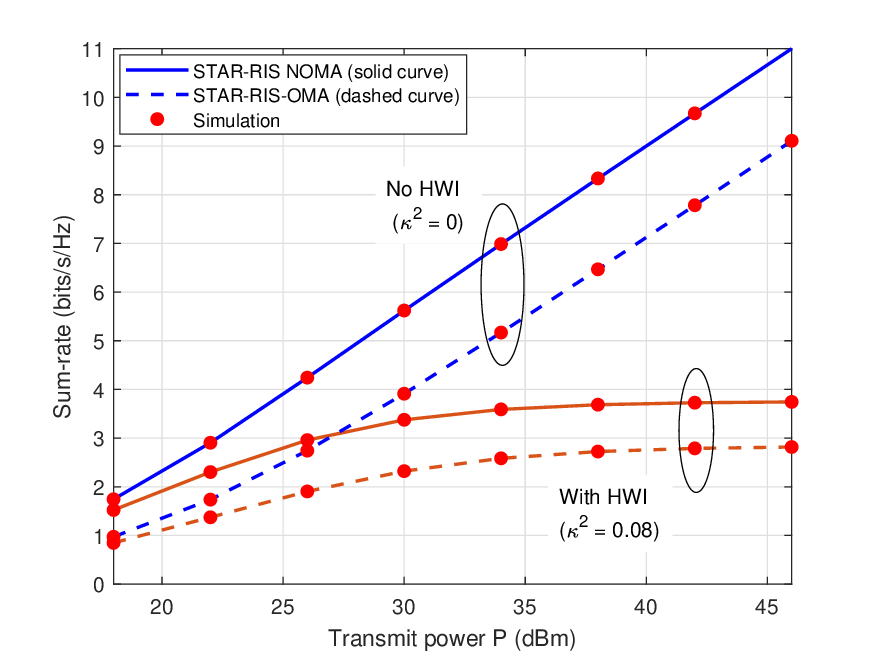}}
	\caption{Sum-rate: NOMA versus OMA.}
	\label{Fig13}
\end{figure}

In Fig. 5, we plot the outage probabilities (OP) at the two users versus the transmit power $P$. The results show that the OP is lower when the transmit power is higher and/or the HWI level is smaller. Also, the indoor user achieves better performance than its counterpart. This is mainly because it is located nearer the STAR-RIS than the outdoor user. In addition, the analysis curves match well with the simulation curves, which demonstrates the accuracy of the derived expressions. The OP under a different fading condition of the channel from the STAR-RIS to the indoor user (i.e., $\alpha_1 = 2$, $\mu_1 = 2$, and $\Omega_1 = 1$) is shown in Fig. 6. Similar observations can be made as in the previous scenario. Moreover, the OP of the indoor user is lower in this scenario compared to the previous one, due to the better channel condition.

The ergodic capacities of the two users versus the transmit power $P$ under different scenarios are shown in Fig. 7 and Fig. 8. It can be observed that the capacities are improved with the increase of the transmit power and/or the decrease of HWI level. Also, at the high SNR regime, the impact of HWI becomes more serious. Specifically, for the indoor user, the capacity increases logarithmically with respect to $P$ when there is no HWI, but it is saturated at the high SNR region when there presents the HWI. For the outdoor user, the capacity is saturated at the high SNR due to the HWI and/or the NOMA detection principle. These agree with the analysis performed in Section IV.B. On the other hand, the result in Fig. 8 reveals that impact of the HWI on the capacity at the low SNR region is marginal. It is also worth noting that the asymptotic expression of the capacity at the low SNR, i.e., Eq. (43), match well with the simulation results.

To further evaluate impacts of the HWI on the system performance, we plot in Fig. 9 the OP and capacities versus the HWI level $\kappa^2$. It can be seen that higher HWI results in the higher OP and the smaller capacities. This makes sense since a larger HWI level further reduces the SIDNR values achieved at both users (cf. Eqs. (4)-(5)). In addition, the results show that the indoor user is more sensitive to the HWI than the outdoor user. This is due to the impacts of the co-channel interference on the two users are different. Specifically, the outdoor user is impacted by the interference from the indoor user, whereas the indoor user enjoys free-interference when perfect SIC is assumed.  

In Fig. 10 and Fig. 11, we plot the OP and capacities achieved in the STAR-RIS with the MS protocol, respectively. In the MS protocol, five RIS elements operates in a reflection mode for signal transmission to the indoor user, whilst the remaining four RIS elements operate in a refraction mode for signal transmission to the outdoor user. We notice that similar observations can be made as in the ES protocol shown in Fig. 5 and Fig. 7. Also, the results reveal that the MS protocol performs worse than the ES protocol in terms of the OP and the capacities. Note that the advantages of the ES protocol over the MS protocol were reported in previous works where a far-field scenario was considered, e.g., \cite{Mu}. Also, the performance improvement of the ES protocol over the MS protocol is achieved at a cost of higher implementation complexity \cite{Mu}.

Finally, Fig. 12 and Fig. 13 compare the OP and the sum-rate between the STAR-RIS-NOMA system and the STAR-RIS OMA system, respectively. In both systems, the target rates of $R_I=R_O=0.5$ (bit per channel use) is used as in \cite{Yue1}. Note that in the STAR-RIS OMA system, the SNR threshold is calculated as $\gamma_{th,\chi}^{OMA}=2^{2R_{\chi}}-1$. Also, the sum-rate is defined as the summation of the rates of the indoor user and the outdoor user. It can be seen from Fig. 12 that the users in STAR-RIS NOMA could achieve lower OP than those in the STAR-RIS OMA counterpart. Also, the STAR-RIS NOMA system could achieve a higher sum-rate than its counterpart for both case of no hardware impairments (i.e., $\kappa^2=0$) and with hardware impairments (i.e., $\kappa^2=0.08$).

\section{Conclusions}
We have analyzed the performance of a STAR-RIS-assisted downlink NOMA THz system with realistic indoor outdoor fading channel models in the presence of residual hardware impairments. The e2e channels based on new statistical results of the weighted sum of $\alpha-\mu$ variates as well as the Gaussian mixture have been characterized. Analytical performance evaluation of the outage probabilities and capacities has been performed. Also, we have compared the achieved performance between two STAR-RIS protocols of energy splitting and mode switching. The future works will be considerations of multiple antennas at transceivers and/or mixed near-field/far-field users on each side of the STAR-RIS panel. Optimization of RIS phase-shifts for improved performance is also worth studying. Furthermore, reducing the passive beamforming complexity and channel estimation overhead, especially when the number of STAR-RIS elements is large, is critical. To this end, it is of interest to investigate the system model with the two-timescale transmission scheme\footnote{The readers are referred to \cite{Zhi50} for more details on scheme.}. Additionally, pointing errors may be a serious issue for the THz transmission. Extending this system model to include pointing errors is, therefore, worth investigating.

\appendices
\section{Exact PDF and CDF of Weighted Sum of $\alpha-\mu$ Variates}\label{Appendix_A}
Let us start by rewriting the $\alpha-\mu$ distribution defined in \eqref{Ch2}-\eqref{Ch3} in terms of the $\alpha$-root mean value, defined as $\hat{x}=(\mathbb{E}\{x^{\alpha}\})^{1/\alpha}$, as \cite{Yac}
\begin{align} \label{A1}
f_{X}(x)=\frac{\alpha\mu^{\mu}}{\hat{x}^{\alpha\mu}\Gamma(\mu)}x^{\alpha\mu-1}e^{-\mu\left(\frac{x}{\hat{x}}\right)^{\alpha}},
\end{align}
\noindent and
\begin{align} \label{A2}
F_{X}(x)=\frac{\gamma\left(\mu,\mu\left(\frac{x}{\hat{x}}\right)^{\alpha}\right)}{\Gamma(\mu)},
\end{align}
\noindent It is obvious from \eqref{Ch2} and \eqref{A1} that $\hat{x}=\frac{\Omega}{\beta}\mu^{1/\alpha}$. Now, let $Y=\sum_{m=1}^M\mathcal{A}_{I,m}X_m$ is a weighted sum of $M$ i.i.d. $\alpha-\mu$ variates $X_m$ with parameters of $(\alpha_1,\mu_1,\hat{X}_m)$, where $\hat{X}_m=\frac{\Omega_1}{\beta_1}\mu_1^{1/\alpha_1}$. Then, we can express $Y=\sum_{m=1}^MY_m$, where $Y_m \triangleq \mathcal{A}_{I,m}X_m$ is an $\alpha-\mu$ variate with the PDF $f_{Y_m}(y)$ defined as in \eqref{A1} and parameters of $(\alpha_1,\mu_1,\hat{Y}_m)$, where $\hat{Y}_m=\mathcal{A}_{I,m}\hat{X}_m$. By taking the Laplace transform of the PDF expression of $Y_m$, we have
\begin{align} \label{A3}
\mathcal{L}\{f_{Y_m}\}(s)=\mathbb{E}\{e^{-sY_m}\}=\int_{0}^{\infty}e^{-sy}f_{Y_m}(y)dy.
\end{align}
\noindent Following the result of \cite[Eq. (35)]{Gar}, we can express
\begin{align} \label{A4}
&\mathcal{L}\{f_{Y_m}\}(s)=\nonumber\\
&\frac{\alpha_1\mu_1^{\mu_1}}{(s\hat{Y}_m)^{\alpha_1\mu_1}\Gamma(\mu_1)}\sum_{i=0}^{\infty} \frac{\Gamma(\alpha_1(i+\mu_1))\left(-\mu_1[1/s\hat{Y}_m]^{\alpha_1}\right)^i}{i!}.
\end{align}

Given that $Y_m, m=1,2,..,M$, are independent random variables, the PDF of $Y$ can be obtained as
\begin{align} \label{A5}
f_Y(y)=f_{Y_1}(y)* f_{Y_2}(y)* \cdots * f_{Y_M}(y),
\end{align}
\noindent where * denotes the convolution. By taking the Laplace transform of \eqref{A5}, we have
\begin{align} \label{A6}
\mathcal{L}\{f_Y\}(s)=&\prod_{m=1}^M \mathcal{L}\{f_{Y_m}\}(s) \nonumber\\
=&\left(\frac{\alpha_1\mu_1^{\mu_1}}{s^{\alpha_1\mu_1}\Gamma(\mu_1)}\right)^M\prod_{m=1}^M\frac{1}{\hat{Y}_m^{\alpha_1\mu_1}}\sum_{i=0}^{\infty} \frac{\aleph_{i,m}s^{-i\alpha_1}}{i!}.
\end{align} 
\noindent where $\aleph_{i,m} \triangleq \Gamma(\alpha_1(i+\mu_1))\left(-\mu_1[1/s\hat{Y}_m]^{\alpha_1}\right)^i$. To obtain the PDF of $Y$, we perform the inverse Laplace of \eqref{A6}, which can be expressed as \cite{Sch}
\begin{align} \label{A7}
f_Y(y)\!=\!\mathcal{L}^{-1}\!\{\mathcal{L}\{f_Y\}(s)\}(y)\!=\!\frac{1}{2\pi j}\! \oint_{\mathcal{C}}\! e^{sy}\mathcal{L}\{f_Y\}(s)ds.
\end{align}
\noindent By substituting \eqref{A6} into \eqref{A7}, we obtain
\begin{align} \label{A8}
f_Y(y)=&\left(\frac{\alpha_1\mu_1^{\mu_1}}{\Gamma(\mu_1)}\right)^M \times \left(\frac{1}{2\pi j}\right)\nonumber\\
&\times \oint_{\mathcal{C}}\frac{e^{sy}}{s^{M\alpha_1\mu_1}}\prod_{m=1}^M\frac{1}{\hat{Y}_m^{\alpha_1\mu_1}}\sum_{i=0}^{\infty} \frac{\aleph_{i,m}s^{-i\alpha_1}}{i!}ds.
\end{align}
\noindent It is important to note that we can express
\begin{align} \label{A9}
\prod_{m=1}^M\frac{1}{\hat{Y}_m^{\alpha_1\mu_1}}\sum_{i=0}^{\infty} \frac{\aleph_{i,m}s^{-i\alpha_1}}{i!}=\sum_{i=0}^{\infty}\delta_i s^{-i\alpha_1},
\end{align}
\noindent where $\delta_i$ is a coefficient defined as in \eqref{E41}, which will be proved later. By substituting \eqref{A9} into \eqref{A8}, we have
\begin{align} \label{A10}
f_Y(y)=\left(\frac{\alpha_1\mu_1^{\mu_1}}{\Gamma(\mu_1)}\right)^M \left(\frac{1}{2\pi j}\right) \oint_{\mathcal{C}}\frac{e^{sy}}{s^{M\alpha_1\mu_1}}\sum_{i=0}^{\infty}\delta_i s^{-i\alpha_1}ds.
\end{align}   
\noindent By following the proof in \cite[Eqs. (45)-(47)]{Gar}, we arrive at the result of \textit{Theorem 2}. 

Let us now determine the expression of $\delta_i$. For notation convenient, we define $\mathcal{B}_{i,m}\triangleq \frac{\aleph_{i,m}}{i!\hat{Y}_m^{\alpha_1\mu_1}}=\frac{(-1)^i\beta_1^{\alpha_1(i+\mu_1)}\Gamma(\alpha_1(i+\mu_1))}{i!\mu_1^{\mu_1}(\mathcal{A}_{I,m}\Omega_1)^{\alpha_1(i+\mu_1)}}$. Then, \eqref{A9} is rewritten as
\begin{align} \label{A11}
\underbrace{\prod_{m=1}^M\sum_{i=0}^{\infty} \mathcal{B}_{i,m} s^{-i\alpha_1}}_{\mathcal{T}_M}=\sum_{i=0}^{\infty}\delta_i s^{-i\alpha_1}.
\end{align}

\noindent \underline{\textit{When M=1}}: In this case, we have $\mathcal{T}_1=\sum_{i=0}^{\infty}\mathcal{B}_{i,1} s^{-i\alpha_1}$. Thus, it is straightforward that
\begin{align} \label{A12}
{}_{1}\delta_i=\mathcal{B}_{i,1}.
\end{align}

\noindent \underline{\textit{When M=2}}: In this case, we have
\begin{align} \label{A13} 
\mathcal{T}_2&=\left(\sum_{i=0}^{\infty}\mathcal{B}_{i,1} s^{-i\alpha_1}\right)\left(\sum_{i=0}^{\infty}\mathcal{B}_{i,2} s^{-i\alpha_1}\right)\nonumber\\
&=\sum_{i=0}^{\infty}\sum_{m_2=0}^i\mathcal{B}_{m_2,2}\mathcal{B}_{i-m_2,1}s^{-i\alpha_1}.
\end{align}
\noindent From \eqref{A11} and \eqref{A13}, we obtain
\begin{align} \label{A14}
{}_{2}\delta_i=\sum_{m_2=0}^i\mathcal{B}_{m_2,2}\mathcal{B}_{i-m_2,1}.
\end{align}

\noindent \underline{\textit{When M=3}}: In this case, we can express
\begin{align} \label{A15} 
\mathcal{T}_3&=\left(\sum_{i=0}^{\infty}\mathcal{B}_{i,1} s^{-i\alpha_1}\right)\left(\sum_{i=0}^{\infty}\mathcal{B}_{i,2} s^{-i\alpha_1}\right)\left(\sum_{i=0}^{\infty}\mathcal{B}_{i,3} s^{-i\alpha_1}\right)\nonumber\\
&=\left(\sum_{i=0}^{\infty}{}_{2}\delta_i s^{-i\alpha_1}\right)\left(\sum_{i=0}^{\infty}\mathcal{B}_{i,3} s^{-i\alpha_1}\right).
\end{align}
\noindent From \eqref{A11} and \eqref{A15}, we obtain
\begin{align} \label{A16}
{}_{3}\delta_i=&\sum_{m_3=0}^i{}_{2}\delta_{i-m_3}\mathcal{B}_{m_3,3}\nonumber\\
=&\sum_{m_3=0}^i\mathcal{B}_{m_3,3}\sum_{m_2=0}^{i-m_3}\mathcal{B}_{m_2,2}\mathcal{B}_{i-m_3-m_2,1}.
\end{align}

\noindent \underline{\textit{When M=4}}: Similarly, we can express
\begin{align} \label{A17}
{}_{4}\delta_i=&\sum_{m_4=0}^i{}_{3}\delta_{i-m_4}\mathcal{B}_{m_4,4}\nonumber\\
=&\sum_{m_4=0}^i\mathcal{B}_{m_4,4}\sum_{m_3=0}^{i-m_4}\mathcal{B}_{m_3,3}\sum_{m_2=0}^{i-m_4-m_3}\mathcal{B}_{m_2,2}\mathcal{B}_{i-m_4-m_3-m_2,1}.
\end{align}

\noindent \underline{\textit{When $ M \geq 5$}}: Following the above procedure, we can generalize to the result in \eqref{E41}. 

\noindent Note that by using \eqref{A12}-\eqref{A17}, we can easily evaluate the value of $\delta_i$ via recursive programming. It is also worth mentioning that when $\mathcal{A}_{I,m}=\mathcal{A}_I, \forall m$, (i.e., a sum of i.n.i.d becomes a sum of i.i.d. $\alpha-\mu$ variates), the result in \textit{Theorem 2} is reduced to the main theorem in \cite{Gar}. This completes the proof.

\section{Approximate PDF and CDF of Weighted Sum of Gaussian Mixture Variates}\label{Appendix_B}
We consider $|H_O|=\sum_{m=1}^M \mathcal{A}_{O,m}|\tilde{h}_{O,m}|$, where $|\tilde{h}_{O,m}|$ follows the GM distribution with parameters of $(N,\varpi_n,\varkappa_n,\eta_n^2)$ and their PDF and CDF are given in \eqref{Ch7} and \eqref{Ch8}, respectively. Note that we can also express $|H_O|=\sum_{m=1}^M |\underline{\tilde{h}}_{O,m}|$, where $|\underline{\tilde{h}}_{O,m}|$ is the GM distribution with parameters of $(N,\varpi_n,\mathcal{A}_{O,m}\varkappa_n,\mathcal{A}_{O,m}^2\eta_n^2)$. In particular, the PDF of $|\underline{\tilde{h}}_{O,m}|$ is given by
\begin{align} \label{B1}
f_{|\underline{\tilde{h}}_{O,m}|}(x)&=\sum_{n}^N\varpi_{n}\frac{1}{\sqrt{2\pi}\mathcal{A}_{O,m}\eta_{n}}e^{-\frac{(x-\mathcal{A}_{O,m}\varkappa_{n})^2}{2\mathcal{A}_{O,m}^2\eta_{n}^2}} \nonumber\\
&=\sum_{n}^N\varpi_{n}\phi(x;\mathcal{A}_{O,m}\varkappa_n,\mathcal{A}_{O,m}^2\eta_n^2),
\end{align} 
\noindent where $\phi(x;u,v^2) \triangleq \frac{1}{\sqrt{2\pi}v}e^{-\frac{(x-u)^2}{2v^2}}$. Since $|\underline{\tilde{h}}_{O,m}|, \forall m$ are independent random variables, the PDF of $|H_O|$ can be expresses as
\begin{align} \label{B2}
&f_{|H_O|}(x)=f_{|\underline{\tilde{h}}_{O,1}|}(x)*f_{|\underline{\tilde{h}}_{O,2}|}(x)* \cdots *f_{|\underline{\tilde{h}}_{O,M}|}(x)\nonumber\\
&=\sum_{n_1=1}^N\sum_{n_2=1}^N\cdots\sum_{n_M=1}^N\prod_{m=1}^M\varpi_{n_m}\left[\phi(x;\mathcal{A}_{O,1}\varkappa_{n_1},\mathcal{A}_{O,1}^2\eta_{n_1}^2)* \right. \nonumber\\
&\left. \phi(x;\mathcal{A}_{O,2}\varkappa_{n_2},\mathcal{A}_{O,2}^2\eta_{n_2}^2)*\cdots*\phi(x;\mathcal{A}_{O,M}\varkappa_{n_M},\mathcal{A}_{O,M}^2\eta_{n_M}^2)\right]\nonumber\\
&=\widetilde{\sum_{\mathbf{n}}}\prod_{m=1}^M\varpi_{n_m}\phi(x;\tilde{\varkappa}_{\mathbf{n}},\tilde{\eta}_{\mathbf{n}}^2),
\end{align}   
\noindent where $\tilde{\varkappa}_{\mathbf{n}} \triangleq \sum_{m=1}^M\mathcal{A}_{O,m}\varkappa_{n_m}$ and $\tilde{\eta}_{\mathbf{n}}^2 \triangleq \sum_{m=1}^M\mathcal{A}_{O,m}^2\eta_{n_m}^2$. Note that the last step is obtained due to the fact that the sum of two independent Gaussian variables is normally distributed where its mean is the sum of the two means, and its variance is the sum of the two variances. It is also worth noting that the moment generating function (MGF) of the weighted sum of Gaussian mixture variates $|H_O|$ is given by
\begin{align} \label{B3}
M_{|H_O|}(t)=\mathbb{E}\{e^{t|H_O|}\}=\widetilde{\sum_{\mathbf{n}}}\prod_{m=1}^M\varpi_{n_m}e^{\tilde{\varkappa}_{\mathbf{n}}+\tilde{\eta}_{\mathbf{n}}^2t^2/2}.
\end{align}  
\noindent Also, based on the $\nu^{th}$ moment of a Gaussian variable \cite{Win}, we can calculate the $\nu^{th}$ moment of $|H_O|$ as
\begin{align} \label{B4}
&\mathbb{E}\{|H_O|^{\nu}\}=\nonumber\\
&\left\{\begin{array}{lcr}
\widetilde{\sum\limits_{\mathbf{n}}}\prod\limits_{m=1}^M\varpi_{n_m}\tilde{\eta}_{\mathbf{n}}^{\nu} 2^{\nu/2} \frac{\Gamma((\nu+1)/2)}{\sqrt{\pi}} {}_{1}F_1\left(-\frac{\nu}{2},\frac{1}{2},-\frac{\tilde{\varkappa}_{\mathbf{n}}^2}{2\tilde{\eta}_{\mathbf{n}}^2}\right),\\  \nu \mbox{\textit{ is even}} \\
\widetilde{\sum\limits_{\mathbf{n}}}\prod\limits_{m=1}^M\varpi_{n_m}\tilde{\varkappa}_{\mathbf{n}}\tilde{\eta}_{\mathbf{n}}^{\nu-1}2^{(\nu+1)/2}\frac{\Gamma(\nu/2+1)}{\sqrt{\pi}} {}_{1}F_1\left(\frac{1-\nu}{2},\frac{3}{2},-\frac{\tilde{\varkappa}_{\mathbf{n}}^2}{2\tilde{\eta}_{\mathbf{n}}^2}\right), \\  \nu \mbox{\textit{ is odd}}.\\
\end{array}\right.
\end{align}
\noindent This completes the proof.

\section{Derivation of Ergodic Capacities}\label{Appendix_C}
\subsection{Indoor User}
To derive a closed-form expression of the capacity, we use the result of the PDF $f_{|H_I|}(x)$ in \textit{Theorem 1}. Similar to many research works on downlink NOMA, e.g., \cite{Zuo}, \cite{Guo}, \cite{Zha1}, and \cite{Li6}, we assume that perfect SIC is achieved. Thus, the PDF of the SDNR is calculated as
\begin{align} \label{AC1}
f_{\gamma_I}(x)&=\frac{\partial F_{\gamma_I}(x)}{\partial x}=\frac{\partial}{\partial x}F_{|H_I|}\left(\sqrt{\frac{N_0x}{P_I-x\kappa^2P}}\right) \nonumber\\
&=f_{|H_I|}\left(\sqrt{\frac{N_0x}{P_I-x\kappa^2P}}\right)\frac{\partial}{\partial x}\left(\sqrt{\frac{N_0x}{P_I-x\kappa^2P}}\right)\nonumber\\
&=\frac{\alpha_{\star}\beta_{\star}^{\alpha_{\star}\mu_{\star}}}{\Omega_{\star}^{\alpha_{\star}\mu_{\star}}\Gamma(\mu_{\star})}\left(\frac{N_0x}{P_I-x\kappa^2P}\right)^{(\alpha_{\star}\mu_{\star}-1)/2}\times\nonumber\\
& e^{-\left(\frac{\beta_{\star}}{\Omega_{\star}}\sqrt{\frac{N_0x}{P_I-x\kappa^2P}}\right)^{\alpha_{\star}}}\frac{1}{2\sqrt{x}}\frac{P_I\sqrt{N_0}}{[P_I-x\kappa^2P]^{3/2}}, x < \frac{P_I}{\kappa^2P}.
\end{align}    
\noindent Now, substituting \eqref{AC1} into \eqref{C1} and then performing some variable changes, we obtain
\begin{align} \label{AC2}
&C_{e,I}=\nonumber\\
&\frac{\alpha_{\star}\beta_{\star}^{\alpha_{\star}\mu_{\star}}}{\Omega_{\star}^{\alpha_{\star}\mu_{\star}}\Gamma(\mu_{\star})}\!\int_{0}^{\infty} \!\!\!y^{\alpha_{\star}\mu_{\star}-1}e^{-\left(\frac{\beta_{\star}y}{\Omega_{\star}}\right)^{\alpha_{\star}}}\!\log_2\left(\!1\!+\!\frac{y^2P_I}{N_0\!+\!y^2\kappa^2P}\!\right)\!dy\nonumber\\
&=\!\frac{\alpha_{\star}}{2\Gamma(\mu_{\star})}\!\int_{0}^{\infty}\!\!\! z^{\frac{\alpha_{\star}\mu_{\star}}{2}-1}e^{-z^{\alpha_{\star}/2}}\log_2\!\left(\!1\!+\!\frac{z\Omega_{\star}^2P_I}{N_0\beta_{\star}^2\!+\!z\Omega_{\star}^2\kappa^2P}\!\right)\!dz\nonumber\\
&=\!\frac{1}{\Gamma(\mu_{\star})}\!\int_{0}^{\infty}\!\!\! u^{\mu_{\star}-1}e^{-u}\log_2\!\left(\!1\!+\!\frac{u^{2/\alpha_{\star}}\Omega_{\star}^2P_I}{N_0\beta_{\star}^2\!+\!u^{2/\alpha_{\star}}\Omega_{\star}^2\kappa^2P}\!\right)\!du.
\end{align}   

\noindent Since an exact evaluation of the integral in \eqref{AC2} is challenging, we adopt the Gauss-Laguerre quadrature for an integral approximation \cite{Abr}. The result in \eqref{C2} is thus obtained. 

\noindent \textit{\textbf{High SNR regime}}: When $\kappa^2 \neq 0$, \eqref{AC2} can be rewritten as
\begin{align} \label{AC3}
C_{e,I, ({\kappa^2 \neq 0})}^{\infty}=\frac{\alpha_{\star}}{2\Gamma(\mu_{\star})}\log_2\left(1+\frac{\rho_I}{\kappa^2}\right)\int_{0}^{\infty}\!\!\! z^{\frac{\alpha_{\star}\mu_{\star}}{2}-1}e^{-z^{\alpha_{\star}/2}}dz.
\end{align} 
\noindent The integral in \eqref{AC3} can be evaluated by a variable change of $u=z^{\alpha_{\star}/2}$. As a result, we obtain
\begin{align} \label{AC4}
C_{e,I, {\kappa^2 \neq 0}}^{\infty}=\log_2\left(1+\frac{\rho_I}{\kappa^2}\right).
\end{align} 

On the other hand, when $\kappa^2=0$, \eqref{AC2} can be rewritten as
\begin{align} \label{AC5}
&C_{e,I, (\kappa^2=0)}=\nonumber\\
&\frac{\alpha_{\star}}{2\Gamma(\mu_{\star})}\int_{0}^{\infty}\!\!\! z^{\frac{\alpha_{\star}\mu_{\star}}{2}-1}e^{-z^{\alpha_{\star}/2}}\log_2\left(1+z\frac{\Omega_{\star}^2P_I}{N_0\beta_{\star}^2}\right)dz.
\end{align}   
\noindent At the high SNR regime (i.e., $\bar{\gamma}_I \triangleq P_I/N_0 \rightarrow 0$), by using the approximation of $\log_2(1+x) \overset{x \rightarrow \infty}{\approx} \log_2x$ and a variable change of $u=z\frac{\Omega_{\star}^2P_I}{N_0\beta_{\star}^2}$, we have
\begin{align} \label{AC6}
&C_{e,I, (\kappa^2=0)}=\nonumber\\
&\frac{\alpha_{\star}}{2\Gamma(\mu_{\star})}\left(\frac{\Omega_{\star}^2P_I}{N_0\beta_{\star}^2}\right)^{-\frac{\alpha_{\star}\mu_{\star}}{2}}\!\!\!\!\int_{0}^{\infty}\!\!\! u^{\frac{\alpha_{\star}\mu_{\star}}{2}-1}e^{-\left(\frac{u N_0\beta_{\star}^2}{\Omega_{\star}^2P_I}\right)^{\alpha_{\star}/2}}\log_2u du\nonumber\\
&\overset{(a)}{=}\frac{1}{\Gamma(\mu_{\star})}\int_{0}^{\infty}v^{\mu_{\star}-1}e^{-v}\left[\frac{2}{\alpha_{\star}}\log_2v+\log_2\left(\frac{\Omega_{\star}^2P_I}{N_0\beta_{\star}^2}\right)\right]dv \nonumber\\
&\overset{(b)}{=}\frac{2\psi(\mu_{\star})}{\alpha_{\star}\ln 2}+\log_2\left(\frac{\Omega_{\star}^2P_I}{N_0\beta_{\star}^2}\right),
\end{align} 
\noindent where the step (\textit{a}) is obtained by changing a variable of $v=\left(\frac{u N_0\beta_{\star}^2}{\Omega_{\star}^2P_I}\right)^{\alpha_{\star}/2}$, and the step (\textit{b}) uses the integral results in \cite [Eq.(3.351.3)]{Gra} and \cite [Eq.(4.352.1)]{Gra}. 

\subsection{Outdoor User}
\subsubsection{Mixture of Gamma}
The capacity of the outdoor user is calculated in a similar way. In case of mixture of gamma model, by using the results in \textit{Theorem 3}, we have
\begin{align} \label{AC7}
&f_{\gamma_O}(x)=\frac{\partial F_{\gamma_O}(x)}{\partial x}=\frac{\partial}{\partial x}F_{|H_O|}\left(\sqrt{\frac{N_0x}{P_O-x(\kappa^2P+P_I)}}\right) \nonumber\\
&=\! f_{|H_O|}\!\left(\!\sqrt{\frac{N_0x}{P_O\!-\!x(\kappa^2P\!+\! P_I)}}\right)\!\frac{\partial}{\partial x}\!\left(\!\sqrt{\frac{N_0x}{P_O\!-\!x(\kappa^2P\!+\!P_I)}}\right)\nonumber\\
&=\widetilde{\sum_{\mathbf{n}}}\tilde{a}_{\mathbf{n}}\left(\sqrt{\frac{N_0x}{P_O-x(\kappa^2P+P_I)}}\right)^{\frac{\tilde{b}_{\mathbf{n}}-1}{2}}e^{-\tilde{c}_{\mathbf{n}}\sqrt{\frac{N_0x}{P_O-x(\kappa^2P+P_I)}}} \nonumber\\
&\times\frac{1}{2\sqrt{x}}\frac{P_O\sqrt{N_0}}{[P_O-x(\kappa^2P+P_I)]^{3/2}}, x < \frac{P_O}{\kappa^2P+P_I}.
\end{align}
\noindent By substituting \eqref{AC7} into \eqref{C1} and performing variable changing, we arrive at
\begin{align} \label{AC8}
&C_{e,O} = \nonumber\\
&\!\widetilde{\sum_{\mathbf{n}}}\tilde{a}_{\mathbf{n}} \int_0^{\infty} y^{\tilde{b}_{\mathbf{n}}-1}e^{-\tilde{c}_{\mathbf{n}}y}\!\log_2\!\!\left(\!\!1\!+\!\frac{y^2P_O}{N_0\!+\! y^2(\kappa^2P\!+\!P_I)}\!\right)dy\!.
\end{align}  
\noindent The result in \eqref{C3} is finally obtained by applying the Gauss-Laguerre approximation to \eqref{AC8}.

\noindent \textit{\textbf{High SNR regime}}: At the high SNR regime, by using the approximation of $\log_2(1+x) \overset{x \rightarrow \infty}{\approx} \log_2x$, we can express
\begin{align} \label{AC9}
C_{e,O}^{\infty} =& \widetilde{\sum_{\mathbf{n}}}\tilde{a}_{\mathbf{n}} \left[\int_0^{\infty} y^{\tilde{b}_{\mathbf{n}}-1}e^{-\tilde{c}_{\mathbf{n}}y}\!\log_2\!\!\left(y^2(\kappa^2\bar{\gamma}\!+\!\bar{\gamma})\!\right)dy \right. \nonumber\\
& -\left.\int_0^{\infty} y^{\tilde{b}_{\mathbf{n}}-1}e^{-\tilde{c}_{\mathbf{n}}y}\!\log_2\!\!\left(y^2(\kappa^2\bar{\gamma}\!+\!\bar{\gamma}_I)\!\right) \right]dy\nonumber\\
=&\log_2\left(\frac{\bar{\gamma}+\kappa^2\bar{\gamma}}{\bar{\gamma}_I+\kappa^2\bar{\gamma}}\right)\widetilde{\sum_{\mathbf{n}}}\tilde{a}_{\mathbf{n}}\int_0^{\infty}y^{\tilde{b}_{\mathbf{n}}-1}e^{-\tilde{c}_{\mathbf{n}}y}dy,
\end{align} 

\noindent On the other hand, it is worth noting from the PDF expression of $|H_O|$ in \eqref{E5} that $\int_0^{\infty}f_{|H_O|}(x)dx=\int_0^{\infty}\widetilde{\sum_{\mathbf{n}}}\tilde{a}_{\mathbf{n}}x^{\tilde{b}_{\mathbf{n}}-1}e^{-\tilde{c}_{\mathbf{n}}x}dx=1$. Thus, we have
\begin{align} \label{AC10}
&C_{e,O}^{\infty} = \log_2\left(\frac{1+\kappa^2}{\rho_I+\kappa^2}\right).
\end{align} 

\subsubsection{Gaussian Mixture}
For the case of Gaussian mixture, similarly, we have
\begin{align} \label{AC11}
f_{\gamma_O}(x)=&\widetilde{\sum\limits_{\mathbf{n}}}\tilde{\varpi}_{\mathbf{n}}\frac{1}{\sqrt{2\pi}\tilde{\eta}_{\mathbf{n}}}e^{-\frac{\left(\sqrt{\frac{N_0x}{P_O-x(\kappa^2P+P_I)}}-\tilde{\varkappa}_{\mathbf{n}}\right)^2}{2\tilde{\eta}_{\mathbf{n}}^2}}\nonumber\\
&\times\frac{1}{2\sqrt{x}}\frac{P_O\sqrt{N_0}}{[P_O-x(\kappa^2P+P_I)]^{3/2}}, x < \frac{P_O}{\kappa^2P+P_I}.
\end{align}
\noindent and, thus, we also obtain
\begin{align} \label{AC12}
C_{e,O} = &\widetilde{\sum\limits_{\mathbf{n}}}\tilde{\varpi}_{\mathbf{n}}\frac{1}{\sqrt{2\pi}\tilde{\eta}_{\mathbf{n}}}\times \nonumber\\
&\int_0^{\infty}e^{-\frac{\left(u-\tilde{\varkappa}_{\mathbf{n}}\right)^2}{2\tilde{\eta}_{\mathbf{n}}^2}}\log_2\left(1+\frac{u^2P_O}{N_0+u^2(\kappa^2P+P_I)}\right)du \nonumber\\
= &\widetilde{\sum\limits_{\mathbf{n}}}\tilde{\varpi}_{\mathbf{n}}\frac{1}{\sqrt{\pi}}e^{-\frac{\tilde{\varkappa}_{\mathbf{n}}^2}{2\tilde{\eta}_{\mathbf{n}}^2}}\int_0^{\infty}e^{-v^2}e^{\frac{\tilde{\varkappa}_{\mathbf{n}}\sqrt{2}v}{\tilde{\eta}_{\mathbf{n}}}}\nonumber\\
&\times \log_2\left(1+\frac{2\tilde{\eta}_{\mathbf{n}}^2P_Ov^2}{N_0+2\tilde{\eta}_{\mathbf{n}}^2v^2(\kappa^2P+P_I)}\right)dv,
\end{align}
\noindent where the last step is obtained via a variable changing. By applying the Gauss-Laguerre approximation, the result of \eqref{C30} follows. 

\noindent \textit{\textbf{High SNR regime}}: At the high SNR regime, we have
\begin{align} \label{AC13}
C_{e,O}^{\infty} &\!=\!\log_2\!\left(\!\frac{\bar{\gamma}\!+\!\kappa^2\bar{\gamma}}{\bar{\gamma}_I\!+\!\kappa^2\bar{\gamma}}\right)\widetilde{\sum_{\mathbf{n}}}\tilde{\varpi}_{\mathbf{n}}\frac{1}{\sqrt{2\pi}\tilde{\eta}_{\mathbf{n}}}\!\int_0^{\infty}\!\! e^{-\frac{\left(u-\tilde{\varkappa}_{\mathbf{n}}\right)^2}{2\tilde{\eta}_{\mathbf{n}}^2}}du\nonumber\\
&=\log_2\left(\frac{1+\kappa^2}{\rho_I+\kappa^2}\right).
\end{align}
\noindent The result of \eqref{C5} is thus obtained. This completes the proof.

\section{Asymptotic Analysis of Capacity at low SNR}
In the low SNR regime, we approximate the capacity via its first and second derivatives with respect to (w.r.t.) the received SNR. In particular, we have \cite{Ver}
\begin{align} \label{D21}
C_{e,\chi}^{0}=\bar{\gamma}_{R,\chi}\dot{C}_{e,\chi}+\frac{1}{2}\bar{\gamma}_{R,\chi}^2\ddot{C}_{e,\chi}+o(\bar{\gamma}_{R,\chi}^2),
\end{align} 
\noindent where $\dot{C}_{e,\chi}$ and $\ddot{C}_{e,\chi}$ denotes the first and the second derivatives of the ergodic capacity w.r.t. the received SNR $\bar{\gamma}_{R,\chi}$. Here, we define $\bar{\gamma}_{R,\chi} \triangleq P G_{\chi}/N_0$ and $G_{\chi} \triangleq \min\{\mathcal{A}_{\chi,m}^2\}, \forall m$. Thus, the SDNR at the users, defined in \eqref{S4} and \eqref{S5}, can be rewritten as
\begin{align} \label{D22}
\gamma_I=\frac{\bar{\gamma}_{R,I}\rho_I|\tilde{H}_I|^2}{\kappa^2\bar{\gamma}_{R,I}|\tilde{H}_I|^2+1}, 
\end{align} 
\noindent and
\begin{align} \label{D220}
\gamma_O=\frac{\bar{\gamma}_{R,O}\rho_O|\tilde{H}_O|^2}{\kappa^2\bar{\gamma}_{R,O}|\tilde{H}_O|^2+\bar{\gamma}_{R,O}\rho_I|\tilde{H}_O|^2+1}.
\end{align}   
\noindent where $|\tilde{H}_{\chi}| \triangleq |H_{\chi}|/\sqrt{G_{\chi}}$. 

For the indoor user, the first derivative of $\dot{C}_{e,I}$ can be evaluated as
\begin{align} \label{D23}
\dot{C}_{e,I} &= \left. \frac{\partial}{\partial \bar{\gamma}_{R,I}} \mathbb{E}\left\{\log_2(1+\gamma_I) \right\}\right|_{\bar{\gamma}_{R,I} \rightarrow 0}\nonumber\\
&=\!\left.\left(\log_2e\right)\frac{\partial}{\partial \bar{\gamma}_{R,I}} \mathbb{E}\left\{\!\ln\!\left(1\!+\!\frac{\bar{\gamma}_{R,I}\rho_I|\tilde{H}_I|^2}{\kappa^2\bar{\gamma}_{R,I}|\tilde{H}_I|^2\!+\!1}\!\right)\!\right\}\!\right|_{\bar{\gamma}_{R,I} \rightarrow 0}\nonumber\\
&=\left(\log_2e\right) \times \nonumber\\
&\left.\mathbb{E}\left\{\!\frac{\rho_I|\tilde{H}_I|^2}{(\kappa^2\bar{\gamma}_{R,I}|\tilde{H}_I|^2\!+\!1)(1\!+\!(\kappa^2\!+\!\rho_I)\bar{\gamma}_{R,I}|\tilde{H}_I|^2)}\!\right\}\!\right|_{\bar{\gamma}_{R,I} \rightarrow 0} \nonumber\\
&= \left(\log_2e\right)\rho_I\mathbb{E}\{|\tilde{H}_I|^2\}.
\end{align}
\noindent Also, the second derivative of $\ddot{C}_{e,I}$ is obtained as
\begin{align} \label{D24}
&\ddot{C}_{e,I}\!=\! \left.\left(\log_2e\right)\frac{\partial^2}{\partial \bar{\gamma}_{R,I}^2} \mathbb{E}\!\left\{\!\ln\!\left(\!1\!+\!\frac{\bar{\gamma}_{R,I}\rho_I|\tilde{H}_I|^2}{\kappa^2\bar{\gamma}_{R,I}|\tilde{H}_I|^2\!+\!1}\!\right)\!\right\}\!\right|_{\bar{\gamma}_{R,I} \rightarrow 0}\nonumber\\
&= \left(\log_2e\right)\mathbb{E}\Big\{-\rho_I|\tilde{H}_I|^2 \times \Big. \nonumber\\
&\left.\left.\frac{[(2\kappa^2+\rho_I)|\tilde{H}_I|^2+2\kappa^2(\kappa^2+\rho_I)\bar{\gamma}_{R,I}|\tilde{H}_I|^4]}{[(\kappa^2\bar{\gamma}_{R,I}|\tilde{H}_I|^2\!+\!1)(1\!+\!(\kappa^2\!+\!\rho_I)\bar{\gamma}_{R,I}|\tilde{H}_I|^2)]^2}\!\right\}\!\right|_{\bar{\gamma}_{R,I} \rightarrow 0} \nonumber\\
&= - \left(\log_2e\right)\rho_I(2\kappa^2+\rho_I)\mathbb{E}\{|\tilde{H}_I|^4\}.
\end{align}

\noindent Similarly, the first and the second derivative of the capacity of the outdoor user are, respectively, given by
\begin{align} \label{D25}
\dot{C}_{e,O} &= \left. \frac{\partial}{\partial \bar{\gamma}_{R,O}} \mathbb{E}\left\{\log_2(1+\gamma_O) \right\}\right|_{\bar{\gamma}_{R,O} \rightarrow 0}\nonumber\\
&= \left(\log_2e\right)\rho_O\mathbb{E}\{|\tilde{H}_O|^2\},
\end{align}
\noindent and
\begin{align} \label{D26}
\ddot{C}_{e,O}= - \left(\log_2e\right)\rho_O(2\kappa^2+\rho_I+1)\mathbb{E}\{|\tilde{H}_O|^4\}.
\end{align}

\noindent By substituting \eqref{D23}-\eqref{D26} into \eqref{D21}, the result in \textit{Theorem 7} follows.

We now calculate the values of $\mathbb{E}\{|\tilde{H}_{\chi}|^2\}$ and $\mathbb{E}\{|\tilde{H}_{\chi}|^4\}$. Recall that $|\tilde{H}_{\chi}| \triangleq |H_{\chi}|/\sqrt{G_{\chi}}$, where $G_{\chi}$ is a constant. Hence, given the distribution of $|H_{\chi}|$, we can obtain the distribution of $|\tilde{H}_{\chi}|$. In particular, for the indoor user, it is noted from \textit{Theorem 1} that $|H_I|$ is approximated as a $\alpha-\mu$ random variable with parameters of $(\alpha_{\star},\mu_{\star},\Omega_{\star})$. Thus, $|\tilde{H}_{I}|$ can be approximated as a $\alpha-\mu$ random variable with parameters of $(\tilde{\alpha}_{\star},\tilde{\mu}_{\star},\tilde{\Omega}_{\star})$, where these parameters are obtained in a similar manner as those of $|H_I|$ excepting that $\mathcal{A}_{I,m}$ is replaced by $\tilde{\mathcal{A}}_{I,m}$. As a result, we can express the $\nu^{th}$ moment as 
\begin{align} \label{D27}
\mathbb{E}\{|\tilde{H}_{I}|^{\nu}\}=\frac{\Gamma^{\nu-1}(\tilde{\mu}_{\star})\Gamma(\tilde{\mu}_{\star}+{\nu}/\tilde{\alpha}_{\star})}{\Gamma^{\nu}(\tilde{\mu}_{\star}+1/\tilde{\alpha}_{\star})}\tilde{\Omega}_{\star}^{\nu}.
\end{align}  
\noindent For the outdoor user, if $|H_O|$ is approximated via a mixture of gamma, we can approximate $|\tilde{H}_{O}|$ via a mixture of gamma with parameters of $(\underline{\tilde{a}}_{\mathbf{n}}, \underline{\tilde{b}}_{\mathbf{n}}, \underline{\tilde{c}}_{\mathbf{n}})$. Note that these parameters are calculated similar to those on \textit{Theorem 3} excepting that $\mathcal{A}_{O,m}$ is replaced by $\tilde{\mathcal{A}}_{O,m}=\mathcal{A}_{O,m}/\sqrt{G_{O}}$. Consequently, the $\nu^{th}$ moment is given by \cite{Le2}
\begin{align} \label{D28}
\mathbb{E}\{|\tilde{H}_{O}|^{\nu}\}=\widetilde{\sum_{\mathbf{n}}}\underline{\tilde{a}}_{\mathbf{n}}\Gamma(\underline{\tilde{b}}_{\mathbf{n}}+\nu)\underline{\tilde{c}}_{\mathbf{n}}^{-(\underline{\tilde{b}}_{\mathbf{n}}+\nu)}.
\end{align}  
\noindent On the other hand, if a Gaussian mixture model is used for $|H_O|$, it is straightforward to show that 
\begin{align} \label{D29}
\mathbb{E}\{|\tilde{H}_{O}|^2\}=\widetilde{\sum_{\mathbf{n}}}\underline{\tilde{\varpi}}_{\mathbf{n}}(\underline{\tilde{\varkappa}}_{\mathbf{n}}^2+\underline{\tilde{\eta}}_{\mathbf{n}}^2)
\end{align} 
\noindent and 
\begin{align} \label{D30}
\mathbb{E}\{|\tilde{H}_{O}|^4\}=\widetilde{\sum_{\mathbf{n}}}\underline{\tilde{\varpi}}_{\mathbf{n}}(\underline{\tilde{\varkappa}}_{\mathbf{n}}^4+6\underline{\tilde{\varkappa}}_{\mathbf{n}}^2\underline{\tilde{\eta}}_{\mathbf{n}}^2+3\underline{\tilde{\eta}}_{\mathbf{n}}^4),
\end{align} 
\noindent where $\underline{\tilde{\varkappa}}_{\mathbf{n}}$ and $\underline{\tilde{\eta}}_{\mathbf{n}}$ are calculated similarly to $\tilde{\varkappa}_{\mathbf{n}}$ and $\tilde{\eta}_{\mathbf{n}}$ in \textit{Theorem 4}, respectively, excepting that $\mathcal{A}_{O,m}$ is replaced by $\tilde{\mathcal{A}}_{O,m}$. This completes the proof.

\end{document}